# The Influence of Global Self-Heating on the Yarkovsky and YORP Effects

B. Rozitis[a] and S. F. Green[a]

[a]*Planetary and Space Sciences, Department of Physical Sciences, The Open University, Walton Hall, Milton Keynes, MK7 6AA, UK*

No. of Manuscript Pages: 39
No. of Figures: 12
No. of Tables: 3

Please direct editorial correspondence and proofs to:

Benjamin Rozitis
Planetary and Space Sciences
Department of Physical Sciences
The Open University
Walton Hall
Milton Keynes
Buckinghamshire
MK7 6AA
UK

Phone: +44 (0) 1908 655169
Fax: +44 (0) 1908 655667

Email: b.rozitis@open.ac.uk

Email address of co-author: s.f.green@open.ac.uk

# ABSTRACT

In addition to collisions and gravitational forces, there is a growing amount of evidence that photon recoil forces from the asymmetric reflection and thermal re-radiation of absorbed sunlight are primary mechanisms that are fundamental to the physical and dynamical evolution of small asteroids. The Yarkovsky effect causes orbital drift, and the Yarkovsky-O'Keefe-Radzievskii-Paddack (YORP) effect causes changes in the rotation rate and pole orientation. We present an adaptation of the Advanced Thermophysical Model to simultaneously predict the Yarkovsky and YORP effects in the presence of global self-heating that occurs within the large concavities of irregularly shaped asteroids, which has been neglected or dismissed in all previous models. It is also combined with rough surface thermal-infrared beaming effects, which have been previously shown to enhance the Yarkovsky-orbital-drift and dampen on average the YORP-rotational-acceleration by orders of several tens of per cent. Tests on all published concave shape models of near-Earth asteroids, and also on one hundred Gaussian-random-spheres, show that the Yarkovsky effect is sensitive to shadowing and global self-heating effects at the few per cent level or less. For simplicity, Yarkovsky models can neglect these effects if the level of accuracy desired is of this order. Unlike the Yarkovsky effect, the YORP effect can be very sensitive to shadowing and global self-heating effects. Its sensitivity increases with decreasing relative strength of the YORP-rotational-acceleration, and doesn't appear to depend greatly on the degree of asteroid concavity. Global self-heating tends to produce a vertical offset in an asteroid's YORP-rotational-acceleration versus obliquity curve which is in opposite direction to that produced by shadowing effects. It also ensures that at least one critical obliquity angle exists at which zero YORP-rotational-acceleration occurs. Global self-heating must be included for accurate predictions of the YORP effect if an asteroid exhibits a large shadowing effect. If global self-heating effects are not included then it is found in ~75 per cent of cases that better predictions are produced when shadowing is also not included. Furthermore, global self-heating has implications for reducing the sensitivity of the YORP effect predictions to detailed variations in an asteroid's shape model.

# 1. INTRODUCTION

## 1.1 The Yarkovsky and YORP Effects

The asteroidal Yarkovsky and Yarkovsky-O'Keefe-Radzievskii-Paddack (YORP) effects are orbital drift and changing spin state, respectively, caused by the asymmetric reflection and thermal re-radiation of sunlight from an irregularly shaped asteroid (see review by Bottke et al. 2006). Fig. 1 shows a schematic of how these two effects arise for an ideal spherical asteroid with two wedges attached at different angles, which also has non-zero surface thermal inertia and is rotating in a prograde sense (copied from Rozitis & Green 2012). As non-zero thermal inertia causes a surface to retain heat, the highest surface temperatures are shifted away from the subsolar point leading to excess thermal emission on the afternoon side of the asteroid. The resulting net photon force pushes in the same direction as the orbital motion and causes the orbit to expand (Yarkovsky effect). The orbit would shrink if the asteroid rotates in a retrograde sense, as the situation would be reversed. This effect is also referred to as the diurnal Yarkovsky effect since its direction and magnitude is dependent on the asteroid rotation. A seasonal Yarkovsky effect also exists for asteroids with non-zero obliquities, which is caused by the alternate heating and delayed thermal emission of the two asteroid hemispheres. The seasonal effect always causes the orbit to shrink, and becomes important for asteroids with very high thermal inertias. As also shown in Fig. 1, photon torques created by reflected sunlight and thermally emitted radiation from the two wedges act in opposite directions about the asteroid centre of mass. Since the wedges are mounted at different angles the reflection and emission directions are also different and the torques do not cancel out. Depending on the shape asymmetry, the resultant torque increases or decreases the rotation rate and can also shift the orientation of the spin axis (YORP effect). Both effects have a number of important implications for the dynamical and physical evolution of small asteroids.

The Yarkovsky effect delivers asteroids smaller than 40 km in size from the main-belt to resonance zones capable of transporting them to Earth-crossing orbits, and dispersing asteroid families. It can make the very close encounters of potentially hazardous asteroids with the Earth very difficult to predict, such as the case of (54509) Apophis (Giorgini et al. 2008; Shor et al. 2012), and adds complications for determining the ages of unbound asteroid pairs (Duddy et al. 2012, 2013). Direct detection of Yarkovsky orbital drift has been achieved by sensitive radar ranging for (6489) Golevka (Chesley et al. 2003), and by deviations from predicted ephemerides over a long time span for (152563) 1992 BF (Vokrouhlický, Chesley & Matson 2008) and for 54 other near-Earth asteroids (Nugent et al. 2012). It has also been indirectly detected through the observed orbital distribution of the Karin cluster asteroid family (Nesvorný & Bottke 2004).

YORP spin-up and spin-down of asteroids smaller than 40 km in size can explain their observed excesses of very fast and slow rotators (Pravec et al. 2008). Spin-up of small rubble pile asteroids (gravitational bound aggregates) can force them to change shape and/or undergo mass shedding (Holsapple 2010), and numerical simulations have demonstrated that continued spin-up can produce binary asteroids (Walsh, Richardson & Michel 2008). Approximately 15 per cent of near-Earth asteroids are inferred to be binaries (Pravec & Harris 2007), and radar observations of binary (66391) 1999 KW4 (Ostro et al. 2006) reveal shapes and orbital properties that are consistent with a formation by continued YORP spin-up. Unbound asteroid pairs have been suggested to have formed from contact-binary asteroids that have undergone YORP-induced rotational fission (Pravec et al. 2010), and the clustering of spin axes observed in asteroid families can be explained by YORP-induced spin axis changes (Vokrouhlický, Nesvorný & Bottke 2003). YORP-rotational-acceleration has been directly detected for asteroids (54509) YORP (Lowry et al. 2007; Taylor et al. 2007), (1862) Apollo (Kaasalainen et al. 2007; Ďurech et al. 2008a), and (1620) Geographos (Ďurech et al. 2008b) by observing very small phase shifts in their rotational light-curves over several years. A fourth probable detection exists for (3103) Eger (Ďurech et al. 2012) which remains to be conclusively confirmed.

**1.2 Modelling the Yarkovsky and YORP Effects**

To accurately predict the Yarkovsky and/or YORP effect acting on an asteroid, any model must take into account the asteroid's size and shape, mass and moment of inertia, surface thermal/reflection/emission properties, rotation state, and its orbit about the Sun. A variety of analytical, numerical, and semi-analytical models have been developed to study these effects, and these models are briefly reviewed in section 1.2 of Rozitis & Green (2012). In this work, a numerical model is developed and used since complex and more accurate physics for highly irregular shapes can be implemented more easily into numerical models than into their analytical counterparts. A typical Yarkovsky/YORP effect numerical model represents the irregular 3D shape of an asteroid by a mesh of triangular facets, and determines temperatures for each facet by non-linear 1D heat conduction whilst sometimes taking into account projected shadows. The photon recoil forces and torques are usually calculated by using the Planck function, the facet normal directions, and a Lambertian scattering/emission model. These are then summed across the asteroid surface, averaged over the asteroid rotation and orbit, and combined with an estimate of the asteroid's mass and moment of inertia to give the orbital drift and spin state change.

Using such models, the Yarkovsky-orbital-drift has been shown to be proportional to the cosine (diurnal effect) or sine (seasonal effect) of the asteroid obliquity, and inversely proportional to the asteroid diameter and bulk density (Bottke et al. 2006). In a complicated way, it is dependent on the heliocentric distance, thermal inertia, rotation period, and Bond albedo, since they are all related to one another via the thermal parameter [see section 3.1 of Rozitis & Green (2012) for an example parameter study]. The Yarkovsky-orbital-drift is also not highly sensitive to subtle variations in an asteroid's shape. In contrast, the YORP-rotational-acceleration is inversely proportional to the asteroid bulk density and to the square of the diameter and heliocentric distance (Rubincam 2000). It is independent of Bond albedo and thermal inertia (Čapek & Vokrouhlický 2004), and is sensitive in a complicated way to the asteroid obliquity and shape (Vokrouhlický & Čapek 2002).

For the current Yarkovsky and YORP effect detected asteroids, their theoretical predictions match the sign and strength of the observed values reasonably well using physical properties that have been inferred by various observational methods. However, light-curve observations of asteroid (25143) Itokawa fail to show a strong YORP-rotational-deceleration (Ďurech et al. 2008a) that is predicted by YORP effect modelling using the Hayabusa-derived shape models (Scheeres et al. 2007). It remains uncertain as to whether this is caused by an unknown non-uniform internal bulk density distribution (Scheeres & Gaskell 2008), or is a product of specific model assumptions and simplifications.

Further investigations into the YORP effect have revealed its predictions to be highly sensitive to unresolved shape features (Statler 2009) and to the shape model resolution (Breiter et al. 2009), such that the error in any prediction could have unity order. In particular, adding additional shape detail appears to add vertical offsets in the YORP-rotational-acceleration versus obliquity prediction curves, which is demonstrated for example in figure 14b of Statler (2009) and figure 1 of Breiter et al. (2009). Again, it is uncertain whether this high sensitivity to subtle shape variations is wholly or partly a product of specific model assumptions and simplifications.

In an attempt to improve the accuracy of Yarkovsky and YORP effect models, our previous work has investigated the influence of rough surface thermal-infrared beaming on these effects using the Advanced Thermophysical Model or ATPM (Rozitis & Green 2011, 2012). Thermal-infrared beaming has the tendency to re-radiate absorbed sunlight back towards the Sun in a non-Lambertian way, and is caused by unresolved surface roughness occurring at scales ranging from the diurnal thermal skin depth up to the resolution of the shape model used. It is the result of two different processes: a rough surface will have elements orientated towards the Sun that become significantly hotter than a flat surface; and multiple scattering of sunlight and re-absorption of emitted thermal radiation between interfacing rough surface elements increases the surface's capability of solar radiation absorption and heat retention [see figure 2 of Rozitis & Green (2012)]. It was found that beaming, on average, enhanced the Yarkovsky-orbital-drift whilst it dampened the YORP-rotational-acceleration by orders of several tens of per cent. The Yarkovsky effect was sensitive to only the average degree of surface roughness, but the YORP effect was sensitive to both the average degree and the spatial distribution of surface roughness.

**1.3 Global Self-Heating**

The implementation of thermal-infrared beaming into the ATPM, described above, had local self-heating (i.e. multiple scattering of sunlight and radiative heat exchange) occurring within hemispherical craters. However, this implementation neglected global self-heating that can occur within large-scale concavities of an irregular shaped asteroid. In fact, all other Yarkovsky and YORP models do not include the effects of global self-heating by either assuming it has a negligible contribution or is too complicated to implement. The Statler (2009) YORP model did include a correction to the emission vector for surface elements whose sky is partly obscured by other parts of the asteroid surface. Since emission towards these other parts of the surface will be absorbed there is no net recoil force in these directions and so they do not contribute to the Yarkovsky and YORP effects. This partly accounts for global self-heating but doesn't consider that the obscuring parts of the surface are heated by the emission they absorb.

When no or 1D heat conduction is assumed, global self-heating provides a mechanism for heat to be transferred laterally across an asteroid surface. Fig. 2 demonstrates how global self-heating could affect the photon recoil force and torque predictions for an asteroid with a large scale concavity and zero thermal inertia. The large-scale concavity results in shadowed areas that occur at certain geometries, where no direct solar flux is received. In the absence of global self-heating, the shadowed areas have zero temperature and no photon recoil force and torque contributions. However, if the shadowed area receives reflected solar flux and emitted thermal radiation from the opposite side of the concavity that is illuminated then it will be heated and radiate thermal radiation of its own. It will now have small photon recoil force and torque components acting in opposite senses to those of the illuminated area.

To investigate how global self-heating affects the Yarkovsky and YORP effects in general, the adaptation of ATPM to make such predictions in the presence of global self-heating, and also in the presence of combined global self-heating and thermal-infrared beaming is described in Section 2. In Section 3, the adapted ATPM is applied to all near-Earth asteroids with a concave shape model and to one hundred synthetic asteroid shapes generated by the Gaussian-random-sphere method to investigate how shape, global self-heating, and the Yarkovsky and YORP predictions are related. In Section 4, the combined effects of global self-heating and thermal-infrared beaming are investigated to see which of the two effects dominates. Further discussion of the results is given in Section 5, and the key results and conclusions are summarised in Section 6.

## 2. YARKOVSKY AND YORP MODELLING

### 2.1 Model Overview

As in Rozitis & Green (2012), the Yarkovsky and YORP effect model presented here is adapted from the ATPM which was developed to calculate the surface temperature distributions and thermal emissions of atmosphereless planetary bodies (Rozitis & Green 2011). Fig. 3 displays a schematic giving an overview of the physics and geometry used in the ATPM. The global shape model of a planetary body is described in terms of the triangular facet formalism, and the unresolved surface roughness of each shape facet is represented by a separate topography model. Any surface roughness representation can be used in the topography model, but hemispherical craters are utilised since they can accurately reproduce the thermal-infrared beaming effects caused by a range of surface roughness morphologies and spatial scales, and are easy to parameterise. The degree of surface roughness for each shape facet is specified by a roughness fraction, $f_R$, that dictates the fraction of the shape facet area represented by the hemispherical crater model, and the remaining fraction, $(1 - f_R)$, represented by a smooth and flat surface. Since each shape facet has an individually assigned roughness fraction it enables different surface roughness distributions to be created across a modelled planetary body's surface.

When evaluating the diurnal temperature variations, both types of facet (shape and roughness) are larger than the diurnal thermal skin depth (~1 cm) so that lateral heat conduction can be neglected and only 1D heat conduction perpendicular and into the surface can be considered. However, only shape facets are considered larger than the seasonal thermal skin depth (~1 to 10 m) for the same approximations to apply when evaluating the seasonal temperature variations. Therefore, the 1D heat conduction equation is solved with a surface boundary condition throughout: an asteroid rotation when evaluating the diurnal temperature variations for both types of facet; and an asteroid orbit when evaluating the seasonal temperature variations for just the shape facets. Roughness facets are then assumed to follow the same seasonal temperature variations as their parent shape facets.

The effects of local and global self-heating are included in the facet surface boundary conditions. In particular, the shape facet surface boundary conditions include direct and multiple scattered solar radiation, shadowing, and re-absorbed thermal radiation from interfacing shape facets (i.e. global self-heating). Likewise, the roughness facet surface boundary conditions include direct and multiple scattered solar radiation, shadowing, and re-absorbed thermal radiation from interfacing roughness facets (i.e. local self-heating), but they also include additional components of multiple scattered solar radiation and re-absorbed thermal radiation from interfacing shape facets (i.e. combined local and global self-heating). Shadowing of both types of facet is determined by standard ray-triangle intersection tests, and radiative heat exchange between interfacing facets is solved by using view-factors.

If a completely smooth surface is assumed then roughness facets can be neglected from the model and only shape facets are iterated on. However, if a non-smooth surface is assumed then the model must iterate on both shape and roughness facets, which it does in a two step procedure. In the two step procedure, shape facets are iterated on first with the effects of global self-heating included until they all reach convergence, and then the roughness facets are iterated on with the effects of both local and global self-heating included until they reach convergence. During the second step, the global self-heating effects on the roughness facets are calculated using the results of the shape facets determined in the first step. Depending on the assumed surface properties and the degree of shape concaveness the model may require up to 1000 revolutions to converge to a solution.

Once the facet illumination fluxes and temperature variations are calculated they are transformed to reflected and thermally emitted photon recoil forces. These forces are directed along vectors that are anti-parallel to the surface normal of facets, or along vectors that take into account the surface normal of facets and directions where reflected/emitted photons are re-absorbed. The total photon force for each shape facet is calculated by weighting the rough and smooth force components by the shape facet's roughness fraction, and is then converted to a photon torque by taking the cross product of the total photon force vector with the shape facet position vector about the asteroid centre of mass. The Yarkovsky-orbital-drift and YORP-rotational-acceleration are then determined by summing the photon forces and torques over all shape facets, averaging over the asteroid rotation and orbit, and by taking into account the asteroid mass, moment of inertia, and pole orientation. Certain aspects of this modelling process are described in more detail in the following subsections.

## 2.2 Thermal Modelling

Following the methodology outlined in Rozitis & Green (2011, 2012), the temperature $T$ for each shape and roughness facet is determined by solving the energy balance equation, which leads to the surface boundary condition:

$$(1-A_B)\left([1-S(\tau)]\psi(\tau)F_{SUN} + F_{SCAT}(\tau)\right) + (1-A_{TH})F_{RAD}(\tau) + \frac{\Gamma}{\sqrt{4\pi P}}\left(\frac{dT}{dz}\right)_{z=0} - \varepsilon\sigma T^4_{z=0} = 0 \quad (1)$$

where $\varepsilon$ is the emissivity, $\sigma$ is the Stefan-Boltzmann constant, $A_B$ is the Bond albedo, $S(\tau)$ indicates whether the facet is shadowed at normalised time $\tau$, $A_{TH}$ is the albedo at themal-infrared wavelengths, $P$ is the rotational (diurnal temperature variation) or orbital (seasonal temperature variation) period, $\Gamma$ is the thermal inertia, and $z$ is the normalised depth below the asteroid surface. $\psi(\tau)$ is a function that returns the cosine of the Sun illumination angle at normalised time $\tau$, and $F_{SUN}$ is the integrated solar flux at the distance of the asteroid. $F_{SCAT}(\tau)$ and $F_{RAD}(\tau)$ are the total multiple scattered sunlight and re-emitted thermal fluxes incident on a facet, respectively, at normalised time $\tau$. The normalised time and depth, $\tau$ and $z$, are related to the actual time and depth, $t$ and $x$, via $\tau = t/P$ and $z = x/l_d$, where $l_d$ is the diurnal or seasonal thermal skin depth.

In the absence of an internal heat source, heat conduction is described by the normalised 1D heat conduction (diffusion) equation

$$\frac{\partial T}{\partial z} = \frac{1}{4\pi}\frac{\partial^2 T}{\partial z^2}, \quad (2)$$

and since the amplitude of subsurface temperature variations decreases exponentially with depth, it implies an internal boundary condition given by

$$\left(\frac{\partial T}{\partial z}\right)_{z\to\infty} \to 0. \quad (3)$$

A finite difference numerical technique is used to solve the problem by equations (1)-(3), and a Newton-Raphson iterative technique is used to solve the surface boundary condition (full details of which are given in Rozitis & Green 2011). Typically, 400 time steps and 40 to 60 depth steps going to a maximum depth of one or two thermal skin depths are used to solve the problem defined here. The terms in equation (1) that are functions of normalised time $\tau$ are more specifically functions of rotational phase when determining the diurnal temperature variation, and functions of orbital phase when determining the seasonal temperature variation. The orbital phase functions are determined by averaging their rotational phase counterparts at each orbital point, which is similar to the approach developed by Vokrouhlický & Farinella (1998) for numerically evaluating the seasonal Yarkovsky effect.

The additional flux contributions from multiple scattered sunlight and re-absorbed thermal emission are calculated using view factors. The view factor from facet $i$ to facet $j$, $f_{i,j}$, is defined as the fraction of the radiative energy leaving facet $i$ which is received by facet $j$ assuming Lambertian emission (Lagerros 1998). It is

$$f_{i,j} = v_{i,j} \frac{\cos\theta_i \cos\theta_j}{\pi d_{i,j}^2} a_j, \qquad (4)$$

where $v_{i,j}$ indicates whether there is line-of-sight visibility between the two facets, $\theta_i$ is facet $i$'s emission angle, $\theta_j$ is facet $j$'s incidence angle, $d_{i,j}$ is the distance separating facet $i$ and $j$, and $a_j$ is the surface area of facet $j$. The view factor given by equation (4) is an approximation since it applies to situations where the separation distance is large relative to the facet area. A more accurate way to calculate view factors in situations where the separation distances are very small is given in Rozitis & Green (2011, 2012). The view factors between two shape facets or between two roughness facets are calculated by either one of these two methods. However, the view factor between a roughness facet and a non-parent shape facet, $f_{i,k,j}$, is calculated by

$$f_{i,k,j} = v_{i,k,j} f_{i,j} \frac{\cos\theta_k}{\cos\theta_i}, \qquad (5)$$

where $k$ is a roughness facet of shape facet $i$, $v_{i,k,j}$ indicates whether there is line-of-sight visibility between roughness facet $k$ and shape facet $j$, and $\theta_k$ is roughness facet $k$'s emission angle. In this case, $f_{i,j}$ is the view factor between shape facets $i$ and $j$, and $\theta_i$ is shape facet $i$'s emission angle. Equation (5) allows roughness-shape view factors to be easily calculated from already existing shape-shape view factors.

Utilising the view factors, the multiple scattered flux leaving shape facet $i$, $G_i(\tau)$, is written as

$$G_i(\tau) = A_B \left( F_{SUN} [1 - S_i(\tau)] \psi_i(\tau) + \sum_{j \neq i} f_{i,j} G_j(\tau) \right), \qquad (6)$$

which can be efficiently solved using a Gauss-Seidel iteration. Similarly, the multiple scattered flux leaving roughness facet $k$, $G_k(\tau)$, is written as

$$G_k(\tau) = A_B \left( F_{SUN} [1 - S_k(\tau)] \psi_k(\tau) + \sum_{l \neq k} f_{k,l} G_l(\tau) + \sum_{j \neq i} f_{i,k,j} G_j(\tau) \right), \qquad (7)$$

where the third term within the brackets acts as a constant during this iteration since it has already been determined in the shape facet iteration given by equation (6). After the Gauss-Seidel iterations have converged to a solution then the total multiple scattered flux incident on a facet is given by

$$F_{SCAT}(\tau) = \frac{G(\tau)}{A_B}. \qquad (8)$$

Only single scattering is considered for thermal emission since planetary surfaces absorb most of the incoming radiation at thermal-infrared wavelengths, i.e. $A_{TH} \sim 0$. For shape facets, the total incident re-emitted thermal flux is calculated using

$$F_{RAD}(\tau) = \varepsilon\sigma \sum_{j \neq i} f_{i,j} T_j^4(\tau), \qquad (9)$$

where $T_j(\tau)$ is the surface temperature of shape facet $j$ at normalised time $\tau$. For roughness facets, it is calculated using

$$F_{RAD}(\tau) = \varepsilon\sigma \sum_{l \neq k} f_{k,l} T_l^4(\tau) + \varepsilon\sigma \sum_{j \neq i} f_{i,k,j} T_j^4(\tau), \qquad (10)$$

where the second term again acts as a constant during the roughness facet iteration since it has already been determined during the first shape facet iteration.

## 2.3 Photon Forces and Torques

There are three types of photons that can impose a recoil force and torque on an asteroid surface: absorbed solar, reflected solar, and thermally radiated. However, only reflected solar and thermally radiated photons are considered because it has been previously shown that absorbed solar photons produce negligible asteroidal orbital perturbations (Žižka & Vokrouhlický 2011) and zero net torque when averaged over the asteroid orbit (Nesvorný & Vokrouhlický 2008; Rubincam & Paddack 2010).

Both reflected solar and thermally radiated photons are assumed to have isotropic (Lambert) emission profiles from a smooth flat facet with a clear view of the sky. This results in a net recoil force anti-parallel to the facet surface normal. However, if other facets are visible to it above its local horizon then photons emitted towards these facets will be re-absorbed resulting in an absorption recoil force that cancels out its emittance recoil force. Taking this into account [see section 2.3 of Rozitis & Green (2012) for a more detailed derivation], the photon recoil force acting on shape facet $i$, $\boldsymbol{p}_i(\tau)$, is given by

$$\boldsymbol{p}_i(\tau) = -\frac{2E_i(\tau)a_i}{3c}\left(\boldsymbol{n}_i(\tau) - \frac{3}{2}\sum_{j\neq i}f_{i,j}\boldsymbol{f}_{i,j}(\tau)\right), \qquad (11)$$

where $a_i$ and $\boldsymbol{n}_i(\tau)$ are the facet area and normal, respectively, and $c$ is the speed of light. $E_i(\tau)$ is the radiant emittance of the facet, which is $G_i(\tau)$ for reflected solar photons and $\varepsilon\sigma T_i^4(\tau)$ for thermally radiated photons. $\boldsymbol{f}_{i,j}(\tau)$ is the unit vector associated with view factor $f_{i,j}$ giving the direction from shape facet $i$ to shape facet $j$. Similarly, the photon recoil force acting on roughness facet $k$, $\boldsymbol{p}_k(\tau)$, is given by

$$\boldsymbol{p}_k(\tau) = -\frac{2E_k(\tau)a_k}{3c}\left(\boldsymbol{n}_k(\tau) - \frac{3}{2}\sum_{l\neq k}f_{k,l}\boldsymbol{f}_{k,l}(\tau) - \frac{3}{2}\sum_{j\neq i}f_{i,k,j}\boldsymbol{f}_{i,j}(\tau)\right). \qquad (12)$$

When transformed into a suitable co-ordinate system [see section 2.3 of Rozitis & Green (2012) for more details], the total smooth and rough surface recoil forces for shape facet $i$, $\boldsymbol{p}_{\text{smooth},i}(\tau)$ and $\boldsymbol{p}_{\text{rough},i}(\tau)$, can be combined as a function of its roughness fraction $f_{\text{R},i}$ to give the total recoil force, $\boldsymbol{p}_{\text{total},i}(\tau)$, as

$$\boldsymbol{p}_{\text{total},i}(\tau) = (1-f_{\text{R},i})\boldsymbol{p}_{\text{smooth},i}(\tau) + f_{\text{R},i}\boldsymbol{p}_{\text{rough},i}(\tau). \qquad (13)$$

The photon torque associated with the total recoil force for shape facet $i$, $\boldsymbol{\varphi}_{\text{total},i}(\tau)$, is then

$$\boldsymbol{\varphi}_{\text{total},i}(\tau) = \boldsymbol{r}_i(\tau) \times \boldsymbol{p}_{\text{total},i}(\tau), \qquad (14)$$

where $\boldsymbol{r}_i(\tau)$ is shape facet $i$'s position vector about the asteroid centre of mass.

## 2.4 Evaluation of the Yarkovsky and YORP Effects

The total photon force and torque acting on an asteroid at a specific point $i$ in its orbit, $\boldsymbol{P}_i$ and $\boldsymbol{\Phi}_i$, can be calculated by summing the recoil forces and torque from each shape facet across the asteroid surface and then by rotation averaging. For determination of the Yarkovsky-orbital-drift, the total force can be split into components that act along: the Sun-asteroid vector, the vector defining the plane of the orbit, and the vector perpendicular to these two. These force vectors have magnitudes $P_{x,i}$, $P_{z,i}$, and $P_{y,i}$ respectively. The orbit-average rate of change in semimajor axis, $da/dt$, for a general orbit is given by

$$\frac{da}{dt} = \frac{2a^2}{P_{\text{ORB}}GM_{\text{SUN}}M_{\text{AST}}}\Delta E, \qquad (15)$$

where $a$ and $P_{\text{ORB}}$ are the semimajor axis and period of the orbit, respectively, $G$ is the gravitational constant, $M_{\text{SUN}}$ is the mass of the Sun, and $M_{\text{AST}}$ is the asteroid mass. $\Delta E$ is the total change in orbital energy integrated over the orbit, and can be found by the summation over $n$ orbital positions:

$$\Delta E = \sum_{i=1}^{n} \Delta t_i \left( v_{x,i} P_{x,i} + v_{y,i} P_{y,i} + v_{z,i} P_{z,i} \right), \tag{16}$$

where $\Delta t_i$ is the time spent at each orbital position, and $v_i$ are the orbital velocity components for the directions defined by the $P_i$ force components (note that $v_z$ is always zero in this geometry).

The YORP-torques can also be transformed into suitable meaningful components. These are the rate of change in angular velocity (rotational acceleration), $d\omega_i/dt$, the rate of change in obliquity, $d\xi_i/dt$, and the precession in longitude, $d\lambda_i/dt$ (Bottke et al. 2006). Using the total torque, $\mathbf{\Phi}_i$, these can be calculated by

$$\frac{d\omega_i}{dt} = \frac{\mathbf{\Phi}_i \cdot \mathbf{d}}{C_\omega}, \tag{17}$$

$$\frac{d\xi_i}{dt} = \frac{\mathbf{\Phi}_i \cdot \mathbf{d}_{\perp 1}}{C_\omega \omega}, \tag{18}$$

$$\frac{d\lambda_i}{dt} = \frac{\mathbf{\Phi}_i \cdot \mathbf{d}_{\perp 2}}{C_\omega \omega}, \tag{19}$$

where $C_\omega$ is the asteroid moment of inertia about its shortest axis, $\omega$ is the angular rotation rate, and $\mathbf{d}$ is the unit vector of the rotation pole direction. $\mathbf{d}_{\perp 1}$ and $\mathbf{d}_{\perp 2}$ are the unit vectors:

$$\mathbf{d}_{\perp 1} = \frac{(\mathbf{o} \cdot \mathbf{d})\mathbf{d} - \mathbf{o}}{\sin \xi}, \tag{20}$$

$$\mathbf{d}_{\perp 2} = \frac{\mathbf{d} \times \mathbf{o}}{\sin \xi}, \tag{21}$$

where $\mathbf{o}$ is the unit vector defining the plane of the asteroid orbit. The orbit-averaged YORP-torque for a general orbit can be determined over $n$ orbital positions using

$$\frac{dY}{dt} = \frac{1}{P_{ORB}} \sum_{i=1}^{n} \Delta t_i \frac{dY_i}{dt}, \tag{22}$$

where $Y$ denotes the three different components of YORP-torque, and $dY_i/dt$ is the YORP-torque strength at orbital position $i$ given by equations (17)-(19).

Alternatively, the YORP-rotational-acceleration for a particular asteroid can be described by a non-dimensional "YORP-coefficient" that is multiplied by a modified solar constant which is then scaled accordingly to the asteroid's size, density, and orbital properties (Rossi, Marzari & Scheeres 2009; Rozitis & Green 2013). The YORP-coefficient, $C_Y$, contains combined information on the asteroid's shape, moment of inertia, and obliquity; and allows the normalised strengths of the YORP effects for different asteroids to be directly compared. In this case, the YORP-rotational-acceleration acting on an asteroid is given by

$$\frac{d\omega}{dt} = \frac{G_1}{a^2 \sqrt{1-e^2} \rho D^2} C_Y, \tag{23}$$

where $G_1$ is the modified solar constant (~6.4 x$10^{16}$ kg m s$^{-2}$), and $\rho$ and $D$ are the asteroid bulk density and diameter respectively.

**2.5 Measuring Shape Concaveness and Degree of Global Self-Heating**

To investigate in general how global self-heating within large concavities of asteroids influences their Yarkovsky and YORP effect predictions, this work utilises two useful measurements of each asteroid shape model. The first useful measurement is the convex volume to concave volume ratio (hereafter referred to as simply the "volume ratio"), which gives an indication of the degree of concavity for a given shape model. This is obtained by fitting a convex hull to the concave shape model to remove all concavities from the asteroid (by utilising a method such as O'Rourke 1998), and then by measuring the volume of both shape model variants and taking their ratio. The convex hull fitting procedure is equivalent to "gift wrapping" the asteroid, and the resulting convex shape model is very similar to those obtained from asteroid light-curve inversion techniques (e.g. Kaasalainen, Torppa & Muinonen 2001). The second useful measurement is the mean total view factor, $<t_{\text{view}}>$, which gives an indication of the overall degree of self-heating that would occur for a given shape or topography model (Rozitis & Green 2011). It is also the mean fraction of sky obscured by other parts of the shape/topography model for any given facet, and for a shape/topography model consisting of $m$ facets it is calculated from the individual facet view factors using

$$\langle t_{\text{view}} \rangle = \frac{1}{m} \sum_{i=1}^{m} \sum_{j \neq i} f_{i,j} . \tag{24}$$

Fig. 4 shows an example total view factor distribution for the radar-derived concave shape model of asteroid (6489) Golevka (Hudson et al. 2000). As indicated, large total view factors occur within deep and large concavities, such as the one located at the south pole, where global self-heating occurs most.

## 2.6 Model Variants

To understand how global self-heating affects the Yarkovsky and YORP effect predictions, six different variants of ATPM with varying complexity are used to try and separate out the influences of each physical process included. The first four model variants assume a smooth surface such that the effects of rough surface thermal-infrared beaming are not included. The first and most basic model variant is the 'convex' model, which uses a convex shape model produced from the concave shape model by convex hull fitting. Since the shape is convex there is automatically no shadowing and global self-heating effects included. This level of model complexity is the same as that used in YORP effect predictions when using light-curve derived shape models (e.g. Kaasalainen et al. 2007; Ďurech et al. 2008a,b; Ďurech et al. 2012). The second model variant is the 'pseudo-convex' model, which uses a concave shape model but both shadowing and global self-heating effects are not included. This level of model complexity is equivalent to that used in some previous YORP effect models (e.g. Scheeres 2007; Scheeres & Gaskell 2008; Steinberg & Sari 2011). The third model variant is the 'shadowing' model, which uses a concave shape model but only shadowing effects are included. Most Yarkovsky and YORP effect models currently in use today have this level of model complexity (e.g. Čapek 2007; Statler 2009; Breiter et al. 2009). The fourth and final smooth model variant is the 'self-heating' model, which uses a concave shape model and both shadowing and global self-heating effects are included. No previous Yarkovsky or YORP effect model has this level of complexity. The final two model variants include rough surfaces such that the effects of rough surface thermal-infrared beaming are included. These are the 'rough shadowing' and the 'rough self-heating' models, and are equivalent to their smooth surface model counterparts, i.e. the 'shadowing' and the 'self-heating' models, that now do include rough surface thermal-infrared beaming effects. To allow direct comparison of the predictions, the 'convex' model variant assumes the same asteroid mass and moment of inertia as that used in the other model variants that utilise the concave shape models. Table 1 summarises these different model variants used.

## 3. GLOBAL SELF-HEATING SENSITIVITY

### 3.1 Test Asteroids

The influence of global self-heating on the Yarkovsky and YORP effect predictions is studied using different concave shape models that are representative of near-Earth asteroid shapes. These include all published radar derived shape models, the spacecraft derived shape models of asteroids (433) Eros and (25143) Itokawa, and one hundred artificial Gaussian-random-sphere shape models [see appendix A of Rozitis & Green (2012) for a method of generation based on Muinonen and Lagerros 1998]. The physical properties of these asteroids and their corresponding shape models are summarised in Table 2. The bulk density quoted for each asteroid is either that measured/assumed in their corresponding shape model paper or it is assumed to be 2500 kg m$^{-3}$. Furthermore, asteroid (4179) Toutatis is assumed to rotate around its long axis rather than in its observed tumbling rotation state.

Some of the interesting physical and shape properties for this selection of asteroids are compared in Fig. 5. Shown in Fig. 5a is the volume ratio plotted as a function of rotation period for the real asteroid shapes. Although the sample size is small, these real asteroid shapes appear to cluster at relatively low volume ratios (~1.02) at rotation periods between 2 and 4 hours. For asteroids with rotation periods outside of this range, the volume ratios appear to be more randomly distributed. Asteroids larger than ~0.15 km in diameter have also been observed to not exceed a critical spin rate of ~2 hours, which suggests that they are low bulk density rubble piles held together by self-gravitation only (Pravec, Harris & Michalowski 2002). Since these low volume ratio asteroids are close to the critical spin rate and have diameters larger than ~0.15 km, then perhaps migration of loose material fills in any large concavities due to their rubble pile nature. This process could be similar to that demonstrated in numerical simulations of binary asteroid formation by YORP spin-up (Walsh, Richardson & Michel 2008).

Fig. 5b shows the mean total view factor for both the real asteroid shapes and the Gaussian-spheres as a function of volume ratio. In general, as the volume ratio increases so does the mean total view factor, which is as expected.

Fig. 5c and Fig. 5d show the frequency distributions for the test asteroid shape models when binned into specific volume ratio and YORP-coefficient ranges. The YORP-coefficient values used here are calculated assuming 0° obliquity and no global self-heating effects. In both figures, the real asteroid shapes and Gaussian-spheres are plotted separately for comparison purposes. The real asteroid shapes occur most frequently at low volume ratios and low YORP-coefficient values, and their frequency appears to drop exponentially at larger values. However, the Gaussian-spheres show different frequency distributions to those of the real asteroid shapes. Most of the Gaussian-spheres fall within a narrow volume ratio range of 1.04 to 1.12, and tend to have higher YORP-coefficient values than the real asteroid shapes. These differences could be caused by the apparent excess of "KW4-like" objects (i.e. highly spherical asteroids with pronounced equatorial ridges) that have low volume ratios and low YORP-coefficient values in the list of real asteroid shapes used [i.e. asteroids (29075) 1950 DA, (66391) 1999 KW4a, (136617) 1994 CC, (276049) 2002 CE26, and 2008 EV5]. It is also possible that the spherical harmonic coefficients derived by Muinonen & Lagerros (1998) for Gaussian-random-sphere shape model generation do not produce shapes that are representative of near-Earth asteroids.

In the rest of this section, global self-heating effects are studied without rough surface thermal-infrared beaming effects included in order to determine their unique influence on the Yarkovsky and YORP effect predictions. The four smooth surface model variants described in Section 2.6 and Table 1 are therefore used; and, for simplicity and like other works, a circular orbit is assumed to remove the dependence of the rotation pole longitude parameter when performing obliquity studies.

## 3.2 Yarkovsky Effect Sensitivity

The Yarkovsky effect sensitivity to global self-heating is studied for two different asteroid thermal inertias at which the diurnal and seasonal orbital drift rates are maximised. The diurnal component is maximised at a thermal inertia of 200 J m$^{-2}$ K$^{-1}$ s$^{-1/2}$, which is also equal to the thermal inertia derived for km-sized near-Earth asteroids by Delbo' et al. (2007). The seasonal component is maximised at a thermal inertia of 2000 J m$^{-2}$ K$^{-1}$ s$^{-1/2}$, which is equivalent to the thermal inertia of a bare rock surface. A Bond albedo of 0.1 is assumed for both the diurnal and seasonal component predictions.

Yarkovsky effect predictions using the four different smooth surface model variants produce orbital drift rates that are very similar to one another and only differ at the few per cent level or less. Shown in Fig. 6 are the percentage differences in orbital drift rate as a function of obliquity for the 'convex', 'pseudo-convex', and 'self-heating' models relative to the 'shadowing' model and averaged across all test asteroids. The error bars on the 'self-heating' model prediction differences represent the one-sigma ranges of difference variations caused by the range of test asteroids used. The morning and afternoon temperature distribution asymmetries that drive the Yarkovsky effect are not significantly affected by global self-heating for the test asteroids used, and hence explains the very low sensitivity of the predicted orbital drift rate to it. However, for the diurnal component (see Fig. 6a), the 'convex', 'pseudo-convex', and 'self-heating' model predictions all show a slight enhancement on average (maximum of ~1.4 per cent) over the 'shadowing' model prediction. This is because all three offer a means to provide a very small fractional addition of radiative input energy from the Sun over that provided in the 'shadowing' model. For the seasonal component (see Fig. 6b), the 'pseudo-convex' predictions are most similar to the 'shadowing' model predictions, and both the 'convex' and 'self-heating' model predictions show a slight enhancement on average (maximum of ~3.6 per cent) in comparison. These very small differences are perhaps more related to the different thermal emission directions between the model variants rather than their differences in input radiative energy. The 'pseudo-convex' and 'shadowing' models utilise the same set of thermal emission directions whilst the 'convex' and 'self-heating' models each have a different set.

## 3.3 YORP Effect Sensitivity

It has previously been demonstrated that the YORP-rotational-acceleration is independent of thermal inertia and Bond albedo (e.g. Čapek & Vokrouhlický 2004), even in the presence of rough surface thermal-infrared beaming (Rozitis & Green 2012). Like the diurnal Yarkovsky effect predictions described previously, fixed values of thermal inertia and Bond albedo of 200 J m$^{-2}$ K$^{-1}$ s$^{-1/2}$ and 0.1 are assumed respectively.

Fig. 7 shows the YORP-rotational-acceleration predictions as a function of obliquity and model variant for all of the real asteroid shapes tested. As indicated, a huge range of sensitivities to the different model variant complexities is apparent, which range from almost no sensitivity [e.g. asteroid (4660) Nereus] to lots of sensitivity [e.g. asteroid (6489) Golevka]. Although not shown, the Gaussian-sphere YORP-rotational-acceleration predictions show a similar sensitivity range. From a qualitative point of view, the 'convex' predictions tend to produce obliquity trends that are similar in shape to the 'pseudo-convex', 'shadowing', and 'self-heating' obliquity trends but have magnitudes that are easily a factor of two or more different. The 'pseudo-convex', 'shadowing', and 'self-heating' obliquity trends are very similar to one another in terms of shape and magnitude but tend to be vertically offset from one another. The vertical offset between the 'self-heating' and 'shadowing' trends tends to occur in the opposite direction to that between the 'shadowing' and 'pseudo-convex' trends. Furthermore, the 'pseudo-convex' critical angles (i.e. the obliquity angles at which zero YORP-rotational-acceleration occurs) for the test asteroids appear to be in better agreement with the 'self-heating' critical angles than the 'shadowing' critical angles.

To quantify the differences exhibited between the four model variant predictions, four types of measurement of their YORP-rotational-acceleration versus obliquity functions (hereafter referred to as simply the "YORP-function") are made for each test asteroid. The first measurement type is the YORP-function difference of the 'convex', 'pseudo-convex', and 'shadowing' predictions relative to the 'self-heating' predictions, which is assumed to be the most correct of the four predictions. It basically gives the ratio of the absolute area bound between the YORP-function of interest and the 'self-heating' YORP-function divided by the absolute area bound by the 'self-heating' YORP-function and the x-axis (i.e. zero YORP-rotational-acceleration at all obliquities), and is given in terms of a percentage. This YORP-function difference, YFD, can be calculated using

$$\text{YFD} = \frac{\int_{\xi=0°}^{\xi=90°} \left| \frac{d\omega}{dt}(\xi) - \frac{d\omega}{dt}(\xi)_{\text{'self-heating'}} \right| d\xi}{\int_{\xi=0°}^{\xi=90°} \left| \frac{d\omega}{dt}(\xi)_{\text{'self-heating'}} \right| d\xi}, \qquad (25)$$

where $d\omega/dt(\xi)$ is the YORP-rotational-acceleration as a function of obliquity. If YFD is greater than 100 per cent then zero YORP-rotational-acceleration at all obliquities provides a better match to the 'self-heating' predictions than the YORP-function of interest. Generally, if YFD is between 50 and 100 per cent then the YORP-function of interest provides a rough match to the 'self-heating' predictions that is better than zero YORP-rotational-acceleration at all obliquities. A reasonable match is obtained for YFD between 10 and 50 per cent, and a good match is obtained for YFD below 10 per cent. Obviously, the predictions are identical if YFD is 0 per cent. The second measurement type gives the fraction, measured in terms of a percentage, of the obliquity space where the 'convex', 'pseudo-convex', and 'shadowing' YORP-functions provide the best match to the 'self-heating' YORP-function for each test asteroid. The third measurement type is the critical angle at which zero YORP-rotational-acceleration occurs for each YORP-function. In some cases there is more than one critical angle [i.e. asteroids (2063) Bacchus, (4769) Castalia, (6489) Golevka, (25143) Itokawa, (29075) 1950 DA, (52760) 1998 ML14, and (66391) 1999 KW4a], and for these cases all of their critical angles are noted and compared. In two other cases [i.e. asteroids (4769) Castalia and (8567) 1996 HW1], there are no critical angles since they have non-zero YORP-rotational-acceleration at all obliquities, and for these cases the obliquity angle which produces the smallest magnitude of YORP-rotational-acceleration is noted and compared. The fourth and final measurement type is the obliquity-averaged vertical offsets between the 'shadowing' and 'pseudo-convex' YORP-functions, <shadowing-pseudo-convex>, and the 'self-heating' and 'shadowing' YORP-functions, <self-heating-shadowing>, measured in terms of the YORP-coefficient. Taking their ratio, <self-heating-shadowing>/<shadowing-pseudo-convex>, allows the magnitude and direction of the vertical offsets to be compared. Table 3 summarises the results of these comparison tests for each asteroid shape investigated.

Shown in Fig. 8 are the YORP-function differences for the 'convex', 'pseudo-convex', and 'shadowing' predictions plotted as functions of volume ratio and 'self-heating' YORP-coefficient for all of the test asteroid shapes used. As indicated, there appears to be no trend with volume ratio but a loose trend seems to exist with YORP-coefficient, i.e. the YORP-function differences decrease with increasing YORP-coefficient. The 'convex' predictions are on average the worst and have many test asteroid shapes with YFD values greater than 100 per cent. The 'pseudo-convex' and 'shadowing' predictions appear to produce similar YFD values but, as indicated in Table 3, the 'pseudo-convex' predictions produce the lowest values, and therefore the best match to the 'self-heating' predictions, for 90 out of the 124 asteroid shapes tested. For comparison, the 'convex' and 'shadowing' predictions produce the best match for only 6 and 28 out of the 124 asteroid shapes tested respectively.

Similarly, as summarised in Table 3, the 'pseudo-convex' predictions produce the best match to the 'self-heating' predictions for the largest fraction of the obliquity space when averaged across all asteroid test shapes, i.e. ~66 per cent. This is much greater than the ~5 per cent for the 'convex' predictions, and more than double the ~29 per cent for the 'shadowing' predictions. In terms of having the largest fraction of the obliquity space for an individual object, the 'pseudo-convex' predictions produce the best match to the 'self-heating' predictions for 94 out of the 124 asteroid shapes tested, which compares against the 5 and 25 asteroid shapes for the 'convex' and 'shadowing' predictions respectively.

Fig. 9 compares the 'convex', 'pseudo-convex', and 'shadowing' derived critical angles against the 'self-heating' derived critical angles for the set of test asteroid shapes used. Again, the 'pseudo-convex' derived critical angles best match those derived from the 'self-heating' model variant by showing the strongest correlation. As summarised in Table 3, the 'pseudo-convex' model variant produces the best match for 93 out of the 124 asteroid shapes tested, which compares against the 13 and 15 asteroid shapes for the 'convex' and 'shadowing' model variants respectively. There were 3 test asteroid shapes [i.e. asteroids (4769) Castalia, (52760) 1998 ML14, and one Gaussian-sphere] in which neither of the 'convex', 'pseudo-convex', and 'shadowing' model variants produced the same number of critical angles as that predicted by the 'self-heating' model variant. Furthermore, for the two cases that previously had no critical angles in their 'shadowing' predictions [i.e. asteroids (4769) Castalia and (8567) 1996 HW1], the vertical offsets induced by global self-heating ensured that they had at least one critical angle in their 'self-heating' predictions.

Finally, Fig. 10 shows the vertical offset ratios, <self-heating-shadowing>/<shadowing-pseudo-convex>, as functions of volume ratio and YORP-coefficient. As shown, and also summarised in Table 3, 114 out of the 124 test asteroid shapes have negative ratios indicating that the global self-heating induced offsets are generally opposite in direction to those induced by shadowing. However, there appears to be no trend with volume ratio or YORP-coefficient, and the vertical offset ratios fall within a range described by a median value with a 1-$\sigma$ spread of $-0.85\ ^{+0.36}/_{-0.55}$ (n.b. a ratio of -1 indicates that they are exactly equal and opposite). These vertical offsets can be explained in terms of the available radiative input energy from which YORP-torque is generated. Including shadowing effects removes some of this available radiative input energy, and causes the vertical offset between the 'shadowing' and 'pseudo-convex' predictions. When global self-heating effects are included then some degree of this energy is put back, and causes a vertical offset of opposite direction between the 'self-heating' and 'shadowing' predictions. The opposite direction nature of the vertical offsets explain why the 'pseudo-convex' predictions are generally a better match to the 'self-heating' predictions than the 'shadowing' predictions.

## 4. COMBINED GLOBAL SELF-HEATING AND THERMAL-INFRARED BEAMING

In Section 3, global self-heating effects were studied without rough surface thermal-infrared beaming effects included. Since the Yarkovsky effect predictions were found to be not very sensitive to global self-heating then it is clear that rough surface thermal-infrared beaming effects would dominate, as they typically enhance the orbital drift rate by several tens of per cent (see Rozitis & Green 2012). However, the YORP-rotational-acceleration predictions have been shown to be very sensitive to both global self-heating and thermal-infrared beaming, and it is initially not clear as to how they would combine. In particular, the YORP-rotational-acceleration is dampened by several tens of per cent on average by thermal-infrared beaming, and is highly sensitive to the spatial distribution of surface roughness.

To investigate how they would combine, four extreme examples of global self-heating and thermal infrared beaming sensitivity are chosen from the real asteroid test shapes shown in Fig. 7. These asteroids include: (1620) Geographos which shows very little global self-heating sensitivity despite its large concavities, (6489) Golevka which is very sensitive to both global self-heating and thermal-infrared beaming, (8567) 1996 HW1 which has non-zero YORP-rotational-acceleration at all obliquities for its 'shadowing' predictions, and (52760) 1998 ML14 which has multiple critical angles. The two rough surface model variants described in Section 2.6 and Table 1, i.e. 'rough shadowing' and 'rough self-heating', are applied to these four test cases. Like the study presented in Rozitis & Green (2012), the degree of roughness is allowed to vary in a patchy way across the asteroid surfaces, and one thousand independent random realisations of the possible surface roughness distributions were made to evaluate the degree of scatter in the predictions.

Fig. 11 shows the YORP-rotational-acceleration predictions as a function of obliquity and model variant for these four test cases. As demonstrated, the 'rough shadowing' and 'rough self-heating' YORP-rotational-acceleration predictions are both on average dampened by the same amount by thermal-infrared beaming for all four test cases and across all obliquity values. Furthermore, they both show the same degree of sensitivity to the potential variation of roughness across the asteroid surface. The main conclusions from Rozitis & Green (2012) regarding the influence of rough surface thermal-infrared beaming on the Yarkovsky and YORP effects are therefore still valid in the presence of global self-heating effects.

## 5. DISCUSSION

As demonstrated in Section 3.2 and Fig. 6, the Yarkovsky effect is not that sensitive to shadowing or global self-heating effects. This is because they do not significantly affect the morning and afternoon temperature distribution asymmetries that drive the Yarkovsky effect. For simplicity, Yarkovsky only prediction models can neglect shadowing or global self-heating effects if the desired level of accuracy required is of the order of a few per cent. Even convex shape models, such as those derived from light-curve inversion (e.g. Kaasalainen, Torppa & Muinonen 2001), are good enough to achieve this level of accuracy. However, since the Yarkovsky effect is highly sensitive to the level of thermal inertia and average degree of surface roughness then effects related to these properties still need to be included in an appropriate model (e.g. Rozitis & Green 2012).

Unlike the Yarkovsky effect, the YORP effect can be very sensitive to shadowing and global self-heating effects, as demonstrated in Section 3.3 and Fig. 7. This is because they change the total radiative input energy available to individual surface elements by different amounts, and therefore affect the way their individual photon torques combine. However, as demonstrated in Fig. 7, not every asteroid shape model with large concavities is affected by it [e.g. asteroid (1620) Geographos]. Also, as demonstrated in Fig. 8, the sensitivity does not appear to depend on the degree of concavity, and it is the magnitude of the YORP-coefficient that is important. For example, asteroids with relatively weak YORP-coefficients are more susceptible to the effects of shadowing and global self-heating. Generally, if there is a large difference between the 'pseudo-convex' and the 'shadowing' YORP predictions then it is likely that global self-heating will induce a large difference too. In these high sensitivity cases, global self-heating should be included to make accurate predictions. However, for the less sensitive cases then the 'pseudo-convex' model variant (no shadowing or global self-heating effects) could be used to make suitable predictions, as it produced a better match to the 'self-heating' model variant in ~75 per cent of cases tested. This is because the YORP-function vertical offsets produced by global self-heating tend to cancel out those generated by shadowing, and the 'self-heating' predictions end up close to the original 'pseudo-convex' predictions. As indicated in Fig. 8 and Table 3, the 'pseudo-convex' predictions are most similar to the 'self-heating' predictions with the 'convex' predictions being the worst match. In Section 4, it was demonstrated that rough surface thermal-infrared beaming in the presence of global self-heating affected the YORP predictions in the same way as described previously in Rozitis & Green (2012). If surface roughness and its thermal-infrared beaming are not explicitly modelled then one can simply assume that it will on average dampen the predicted YORP-rotational-acceleration by several tens of per cent, and add an additional uncertainty of a similar size due to potential variations in roughness across the surface [see figure 20 of Rozitis & Green (2012)].

The YORP-function vertical offsets produced by shadowing and global self-heating effects are similar in magnitude to those seen in figure 14b of Statler (2009) and figure 1 of Breiter et al. (2009). In these works, the vertical offsets were caused by adding more detail to the shape models of test Gaussian-spheres and that of asteroid (25143) Itokawa, and grew larger with increasing shape detail. When more shape detail is added it increases the number of projected shadows that can be generated, and also increases the degree of self-heating that the global shape can undergo. Since these works did not consider global self-heating effects and only shadowing ones, then the increasing amount of radiative input energy lost by shadowing is potentially not being put back by global self-heating. If this is the case then the increasing vertical offsets caused by shadowing would be reduced or cancelled out by increasing vertical offsets of opposite direction by global self-heating, which also ensure that at least one critical angle exists. Once global self-heating has been accounted for then the different shape detail YORP-functions would more or less overlap, and therefore reduce the detailed shape sensitivity of the YORP effect predictions. The end result would be similar to that shown in figure 1 of Scheeres & Gaskell (2008), which shows overlapping 'pseudo-convex' YORP-functions for different resolutions of the (25143) Itokawa shape model. The variation between the YORP-functions in this case is a lot less than the offset YORP-functions shown in figure 1 of Breiter et al. (2009). However, even when accounting for any global self-heating offsets it still appears that a YORP-rotational-deceleration should have been detected for (25143) Itokawa by now (Ďurech et al. 2008a), and a non-uniform internal bulk density distribution would be the likely cause for its non-detection (Scheeres & Gaskell 2008). Repeating the studies of Statler (2009) and Breiter et al. (2009) with global self-heating effects included should verify this and is a subject for future work. However, Fig. 7 already demonstrates that the 'self-heating' prediction is very similar to the 'pseudo-convex' prediction for the lowest resolution spacecraft-derived shape model of (25143) Itokawa.

Related to the YORP shape sensitivity problem, the step from a convex shape model to a concave one is an extreme example of more detail being added to an existing shape model. Fig. 8 demonstrates that for large YORP-coefficients, the 'convex' predictions can be very similar to the 'self-heating' predictions that use the concave shape model for some test case asteroids [e.g. asteroid (4660) Nereus]. This presumably happens when the large-scale global shape asymmetries that cause the majority of the YORP-torque are still retained in the detail of a convex shape model. For an asteroid with a measured YORP-rotational-acceleration, its YORP-coefficient can be determined using knowledge of its orbital properties, a measurement of its diameter, and an assumption of its bulk density (see equation 23). Its measured YORP-coefficient can then be used to estimate how accurate a YORP prediction made from its light-curve-derived convex shape model would be by comparing it to the test asteroids displayed in Fig. 8. One of the current YORP effect detected asteroids, (1862) Apollo, has a large YORP-coefficient value of (1.9 ± 0.4) x$10^{-2}$ (Rozitis & Green 2013), which Fig. 8 indicates could have a YORP-function with good accuracy. This could explain why Ďurech et al. (2008a) obtain a light-curve-derived convex shape model that produces a YORP-rotational-acceleration prediction that agrees well with the measured value for realistic physical properties.

# 6. SUMMARY AND CONCLUSIONS

The ATPM presented in Rozitis & Green (2011) has been adapted to simultaneously predict the Yarkovsky and YORP effects acting on an asteroid with the effects of global self-heating included. It has also been combined with rough surface thermal-infrared beaming, which was investigated in Rozitis & Green (2012). This is the first such model of its kind, and a detailed investigation into the influence of shadowing and global self-heating effects on all published concave shape models of near-Earth asteroids, and also on one hundred artificial Gaussian-random-spheres, was performed.

The Yarkovsky effect was found not to be highly sensitive to shadowing or global self-heating effects, as they only affected the orbital drift predictions by a few per cent or less. For simplicity, these effects can be neglected from Yarkovsky only prediction models if the desired level of accuracy is of this order. Furthermore, convex shape models, such as those derived from light-curve inversion (e.g. Kaasalainen, Torppa & Muinonen 2001), are also sufficient to achieve this level of accuracy. However, the effects of rough surface thermal-infrared beaming must still be included, which typically enhance the orbital drift by several tens of per cent (Rozitis & Green 2012).

Unlike the Yarkovsky effect, the YORP effect can be very sensitive to shadowing and global self-heating effects, and the sensitivity appears to depend on the relative strength of the YORP-rotational-acceleration (i.e. weaker relative strengths are more sensitive) rather than the degree of concavity. Global self-heating tends to produce a vertical offset in an asteroid's YORP-rotational-acceleration versus obliquity curve that is opposite in direction and roughly equal in magnitude to that produced by shadowing, and ensures that at least one critical angle exists at which zero YORP-rotational-acceleration occurs. The net result is that a YORP prediction that includes both shadowing and global self-heating effects sometimes ends up being quite similar to a prediction that doesn't include either effect, i.e. a 'pseudo-convex' model. Indeed, in ~75 per cent of cases tested, a model that neglects both shadowing and global self-heating effects produced a better match to the predictions of a model that includes both effects rather than a model that includes only shadowing. A true 'convex' model still produces a rather inaccurate prediction except for some asteroids that have relatively large YORP-rotational-acceleration values. A simple way to assess whether global self-heating should be included in a prediction for a specific asteroid is to see whether there is a large difference between predictions that do and do not include shadowing. The effects of rough surface thermal-infrared beaming when combined with global self-heating still affect the predictions in the same way as described in Rozitis & Green (2012). In particular, the YORP-rotational-acceleration is on average dampened by several tens of per cent, and is highly sensitive to the spatial distribution of surface roughness.

Finally, the vertical offsets caused by shadowing and global self-heating are similar to those seen in studies of the shape sensitivity of the YORP effect, e.g. figure 14b of Statler (2009) and figure 1 of Breiter et al. (2009). Since these studies only included shadowing and no global self-heating effects then the implication from this work is that the predictions using different shape model variants of the same asteroid could more or less overlap once global self-heating has been included, e.g. like figure 1 of Scheeres & Gaskell (2008) which was produced using a 'pseudo-convex' model. If this is the case then this would reduce the overall shape sensitivity of the YORP effect, and perhaps make it possible to make predictions with realistic uncertainties for asteroids with very detailed shape models (such as those obtained from spacecraft or from very high resolution radar observations). However, this remains to be confirmed, and the detailed shape sensitivity of the YORP effect in the presence of global self-heating and rough surface thermal-infrared beaming effects will be studied in detail in a future paper.

**Acknowledgments**

We are grateful to the anonymous referee for several suggested refinements to the manuscript. The work of BR is supported by the UK Science and Technology Facilities Council (STFC).

## APPENDIX A: ROZITIS & GREEN (2012) ERRATA

### A.1 Better View Factor Unit Vector

The better view factor unit vector $\boldsymbol{f}_{i,j}(\tau)$ given in equation (15) of Rozitis & Green (2012) is currently not normalised and needs to be divided by the total view factor $f_{i,j}$ to give it unit length.

### A.2 Figure 7c

Unfortunately, the seasonal Yarkovsky effect acting on asteroid (1620) Geographos displayed in figure 7c of Rozitis & Green (2012) was calculated incorrectly. The miss-calculation affected the shape and magnitude of the prediction with obliquity, and the corrected plot is shown in Fig. A1. This was produced using its light-curve-derived convex shape model (Ďurech et al. 2008b), and by assuming a Bond albedo of 0.1, thermal inertia values of 200 and 2000 J m$^{-2}$ K$^{-1}$ s$^{-1/2}$ for the diurnal and seasonal components respectively, and a smooth surface.

**Tables**

Table 1: Summary of ATPM model variants.

| Model Variant | Shape Type | Shadowing Effects | Global Self-Heating Effects | Rough Surface Thermal-Infrared Beaming Effects | Example Equivalent Model(s) |
|---|---|---|---|---|---|
| 'Convex' | Convex | N/A | N/A | No | Kaasalainen et al. 2007<br>Ďurech et al. 2008a,b<br>Ďurech et al. 2012 |
| 'Pseudo-Convex' | Concave | No | No | No | Scheeres 2007<br>Scheeres & Gaskell 2008<br>Steinberg & Sari 2011 |
| 'Shadowing' | Concave | Yes | No | No | Čapek 2007<br>Statler 2009<br>Breiter et al. 2009 |
| 'Self-Heating' | Concave | Yes | Yes | No | N/A |
| 'Rough Shadowing' | Concave | Yes | No | Yes | Rozitis & Green 2012 |
| 'Rough Self-Heating' | Concave | Yes | Yes | Yes | N/A |

Table 2: Summary of test asteroid physical and shape properties.

| Asteroid | Shape Model Reference | Number of Vertices | Number of Concave Facets | Number of Convex Facets | Convex Volume / Concave Volume | Mean Total View Factor (x0.01) | Diameter of Equivalent Volume Sphere (km) | Bulk Density (kg m$^{-3}$) | Mass (kg) | Moment of Inertia (kg m$^2$) | Semimajor Axis (AU) | Rotation Period (hr) | Yarkovsky Semimajor Axis Drift* (m yr$^{-1}$) | YORP-Coefficient* (x0.01) |
|---|---|---|---|---|---|---|---|---|---|---|---|---|---|---|
| (433) Eros | Thomas et al. 2002 | 3897 | 7790 | 2204 | 1.189 | 0.820 | 16.849 | 2670 | 6.7 x10$^{15}$ | 5.0 x10$^{23}$ | 1.458 | 5.270 | 3 | 1.736 |
| (1580) Betulia | Magri et al. 2007 | 1148 | 2292 | 1246 | 1.016 | 0.127 | 5.390 | 2000 | 1.6 x10$^{14}$ | 6.4 x10$^{20}$ | 2.196 | 6.132 | 7 | 1.611 |
| (1620) Geographos | Hudson & Ostro 1999 | 2048 | 4092 | 854 | 1.124 | 1.167 | 2.568 | 2500 | 2.2 x10$^{13}$ | 3.1 x10$^{19}$ | 1.246 | 5.223 | 21 | 5.470 |
| (2063) Bacchus | Benner et al. 1999 | 256 | 508 | 224 | 1.084 | 0.270 | 0.631 | 2500 | 3.3 x10$^{11}$ | 2.5 x10$^{16}$ | 1.078 | 14.904 | 78 | 0.138 |
| (2100) Rashalom | Shepard et al. 2008 | 1148 | 2292 | 586 | 1.043 | 0.798 | 2.280 | 2400 | 1.5 x10$^{13}$ | 1.0 x10$^{19}$ | 0.832 | 19.797 | 18 | 0.331 |
| (4179) Toutatis | Hudson & Ostro 1995 | 1600 | 3196 | 630 | 1.135 | 1.007 | 2.447 | 2500 | 1.9 x10$^{13}$ | 7.1 x10$^{18}$ | 2.529 | 129.840 | 21 | 2.251 |
| (4486) Mithra | Brozovic et al. 2010 | 3000 | 5996 | 726 | 1.302 | 4.654 | 1.691 | 2000 | 5.1 x10$^{12}$ | 2.6 x10$^{18}$ | 2.202 | 67.500 | 30 | 4.564 |
| (4660) Nereus | Brozovic et al. 2009 | 1148 | 2292 | 1116 | 1.009 | 0.056 | 0.333 | 2000 | 3.9 x10$^{10}$ | 6.7 x10$^{14}$ | 1.489 | 15.160 | 146 | 1.249 |
| (4769) Castalia | Hudson & Ostro 1994 | 2048 | 4092 | 1810 | 1.086 | 0.947 | 1.084 | 2500 | 1.7 x10$^{12}$ | 3.4 x10$^{17}$ | 1.063 | 4.095 | 48 | 0.362 |
| (6489) Golevka | Hudson et al. 2000 | 2048 | 4092 | 610 | 1.162 | 3.688 | 0.530 | 2700 | 2.1 x10$^{11}$ | 7.3 x10$^{15}$ | 2.498 | 6.026 | 60 | 0.168 |
| (8567) 1996 HW1 | Magri et al. 2011 | 1392 | 2780 | 726 | 1.262 | 3.277 | 2.023 | 2000 | 8.7 x10$^{12}$ | 9.6 x10$^{18}$ | 2.046 | 8.762 | 25 | 2.322 |
| (10115) 1992 SK | Busch et al. 2006 | 510 | 1016 | 512 | 1.032 | 0.266 | 1.005 | 2300 | 1.2 x10$^{12}$ | 1.5 x10$^{17}$ | 1.249 | 7.318 | 54 | 1.851 |
| (25143) Itokawa | Gaskell et al. 2006 | 25350 | 49152 | 2614 | 1.132 | 1.548 | 0.327 | 1950 | 3.6 x10$^{10}$ | 7.8 x10$^{14}$ | 1.324 | 12.132 | 178 | 0.408 |
| (29075) 1950 DA prograde | Busch et al. 2007 | 1020 | 2036 | 666 | 1.031 | 0.201 | 1.161 | 3000 | 2.5 x10$^{12}$ | 3.4 x10$^{17}$ | 1.699 | 2.122 | 28 | 0.925 |
| (29075) 1950 DA retrograde | Busch et al. 2007 | 510 | 1016 | 552 | 1.014 | 0.050 | 1.298 | 3500 | 4.0 x10$^{12}$ | 8.4 x10$^{17}$ | 1.699 | 2.122 | 17 | 0.065 |
| (33342) 1998 WT24 | Busch et al. 2008 | 4000 | 7996 | 1924 | 1.035 | 0.279 | 0.415 | 3000 | 1.1 x10$^{11}$ | 2.2 x10$^{15}$ | 0.718 | 3.697 | 124 | 2.196 |
| (52760) 1998 ML14 | Ostro et al. 2001 | 512 | 1020 | 388 | 1.060 | 1.699 | 0.992 | 2500 | 1.3 x10$^{12}$ | 1.3 x10$^{17}$ | 2.412 | 14.980 | 39 | 0.319 |

*Calculated using 0° obliquity, a thermal inertia of 200 J m$^{-2}$ K$^{-1}$ s$^{-1/2}$, a Bond albedo of 0.1, and assuming no global self-heating effects.

Table 2 (continued): Summary of test asteroid physical and shape properties.

| Asteroid | Shape Model Reference | Number of Vertices | Number of Concave Facets | Number of Convex Facets | Convex Volume / Concave Volume | Mean Total View Factor (x0.01) | Diameter of Equivalent Volume Sphere (km) | Bulk Density (kg m$^{-3}$) | Mass (kg) | Moment of Inertia (kg m$^2$) | Semimajor Axis (AU) | Rotation Period (hr) | Yarkovsky Semimajor Axis Drift* (m yr$^{-1}$) | YORP-Coefficient* (x0.01) |
|---|---|---|---|---|---|---|---|---|---|---|---|---|---|---|
| (54509) YORP | Taylor et al. 2007 | 288 | 572 | 190 | 1.165 | 3.367 | 0.113 | 2500 | 1.9 x10$^9$ | 3.0 x10$^{12}$ | 1.006 | 0.203 | 368 | 3.521 |
| (66391) 1999 KW4a | Ostro et al. 2006 | 4586 | 9168 | 1576 | 1.030 | 0.601 | 1.317 | 1970 | 2.4 x10$^{12}$ | 4.6 x10$^{17}$ | 0.642 | 2.765 | 59 | 0.058 |
| (66391) 1999 KW4b | Ostro et al. 2006 | 1148 | 2292 | 1648 | 1.005 | 0.021 | 0.451 | 2810 | 1.3 x10$^{11}$ | 3.8 x10$^{15}$ | 0.642 | 17.422 | 70 | 0.134 |
| (136617) 1994 CC | Brozović et al. 2011 | 2000 | 3996 | 1148 | 1.020 | 0.289 | 0.620 | 2100 | 2.6 x10$^{11}$ | 1.0 x10$^{16}$ | 1.638 | 2.389 | 77 | 0.045 |
| (276049) 2002 CE26 | Shepard et al. 2006 | 1148 | 2292 | 1490 | 1.007 | 0.086 | 3.459 | 900 | 2.0 x10$^{13}$ | 2.5 x10$^{19}$ | 2.233 | 3.293 | 23 | 0.188 |
| 1998 KY26 | Ostro et al. 1999 | 2048 | 4092 | 2294 | 1.026 | 0.269 | 0.026 | 2500 | 2.4 x10$^7$ | 1.7 x10$^9$ | 1.232 | 0.178 | 1134 | 1.452 |
| 2008 EV5 | Busch et al. 2011 | 2000 | 3996 | 924 | 1.032 | 1.079 | 0.405 | 3000 | 1.0 x10$^{11}$ | 1.8 x10$^{15}$ | 0.958 | 3.725 | 120 | 0.276 |
| Gaussian-Spheres | Rozitis & Green 2012 | 578 | 1152 | 268-504 | 1.043-1.236 | 0.452-3.711 | 1.000 | 2500 | 1.3 x10$^{12}$ | (1.6-3.1) x10$^{17}$ | 1.000 | 6.000 | 42-59 | 0.051-7.700 |

*Calculated using 0° obliquity, a thermal inertia of 200 J m$^{-2}$ K$^{-1}$ s$^{-1/2}$, a Bond albedo of 0.1, and assuming no global self-heating effects.

Table 3: Comparison of YORP effect predictions made by the different ATPM model variants for the set of test asteroid shapes used.

| Asteroid | YORP Function Difference (%) | | | Best Obliquity Match (%) | | | Critical Angles (°) | | | | Vertical YORP-Coefficient Offsets (x0.01) | | |
|---|---|---|---|---|---|---|---|---|---|---|---|---|---|
| | 'Convex' | 'Pseudo-Convex' | 'Shadowing' | 'Convex' | 'Pseudo-Convex' | 'Shadowing' | 'Convex' | 'Pseudo-Convex' | 'Shadowing' | 'Self-Heating' | <Shadowing - Pseudo-Convex> | <Self-Heating - Shadowing> | Ratio |
| (433) Eros | 61.0 | 2.52 | 27.9 | 0 | 100 | 0 | 53.8 | 58.1 | 63.5 | 58.5 | 0.224 | -0.231 | -1.033 |
| (1580) Betulia | 32.6 | 4.61 | 9.54 | 0 | 80 | 20 | 58.9 | 58.3 | 59.7 | 58.0 | 0.101 | -0.079 | -0.774 |
| (1620) Geographos | 42.7 | 3.99 | 4.32 | 10 | 40 | 50 | 50.0 | 54.5 | 53.9 | 54.6 | 0.067 | -0.128 | -1.912 |
| (2063) Bacchus | 597 | 5.46 | 47.1 | 0 | 100 | 0 | 56.0 | 18.4, 63.2 | 13.0, 65.3 | 17.8, 62.9 | 0.068 | -0.073 | -1.072 |
| (2100) Rashalom | 48.3 | 6.99 | 31.5 | 0 | 100 | 0 | 56.2 | 50.3 | 44.1 | 52.3 | -0.041 | 0.049 | -1.179 |
| (4179) Toutatis | 44.7 | 4.27 | 0.816 | 0 | 0 | 100 | 62.1 | 59.8 | 60.7 | 60.8 | 0.023 | 0.009 | 0.379 |
| (4486) Mithra | 51.0 | 20.1 | 37.9 | 0 | 80 | 20 | 53.6 | 53.6 | 60.6 | 53.2 | -0.936 | 0.716 | -0.764 |
| (4660) Nereus | 7.48 | 0.645 | 4.41 | 0 | 100 | 0 | 56.2 | 56.1 | 55.2 | 56.2 | -0.027 | 0.028 | -1.035 |
| (4769) Castalia | 944 | 153 | 473 | 10 | 90 | 0 | 58.8 | 56.3 | None (40.0*) | 15.1, 61.4 | 0.331 | -0.289 | -0.872 |
| (6489) Golevka | 448 | 100 | 96.3 | 0 | 70 | 30 | 54.9 | 32.1, 66.9 | 10.6 | 48.4 | 0.172 | -0.305 | -1.776 |
| (8567) 1996 HW1 | 58.9 | 72.0 | 148 | 60 | 20 | 20 | 61.6 | 61.5 | None (90.0*) | 51.2 | -1.278 | 1.027 | -0.803 |
| (10115) 1992 SK | 41.8 | 1.22 | 2.56 | 0 | 100 | 0 | 56.9 | 55.7 | 55.4 | 55.9 | -0.019 | 0.026 | -1.396 |
| (25143) Itokawa | 73.1 | 15.1 | 33.7 | 30 | 40 | 30 | 24.2, 65.4 | 62.2 | 66.9 | 61.7 | -0.159 | 0.135 | -0.848 |
| (29075) 1950 DA prograde | 85.1 | 0.837 | 7.74 | 0 | 100 | 0 | 43.2, 76.2 | 55.5 | 54.2 | 55.5 | -0.042 | 0.042 | -0.993 |
| (29075) 1950 DA retrograde | 65.7 | 1.40 | 3.60 | 0 | 100 | 0 | 60.5 | 23.1, 62.8 | 23.6, 62.6 | 23.2, 62.9 | -0.003 | 0.003 | -1.064 |
| (33342) 1998 WT24 | 46.3 | 1.64 | 4.02 | 10 | 50 | 40 | 54.6 | 53.4 | 51.6 | 52.9 | -0.043 | 0.039 | -0.917 |
| (52760) 1998 ML14 | 128 | 145 | 131 | 20 | 30 | 50 | 65.6 | 65.3 | 66.8 | 25.8, 46.5, 70.3 | -0.003 | -0.051 | 14.655 |

*Obliquity which produces the smallest magnitude of YORP-rotational-acceleration.

Table 3 (continued): Comparison of YORP effect predictions made by the different ATPM model variants for the set of test asteroid shapes used.

| Asteroid | YORP Function Difference (%) | | | Best Obliquity Match (%) | | | Critical Angles (°) | | | | Vertical YORP-Coefficient Offsets (x0.01) | | |
|---|---|---|---|---|---|---|---|---|---|---|---|---|---|
| | 'Convex' | 'Pseudo-Convex' | 'Shadowing' | 'Convex' | 'Pseudo-Convex' | 'Shadowing' | 'Convex' | 'Pseudo-Convex' | 'Shadowing' | 'Self-Heating' | <Shadowing - Pseudo-Convex> | <Self-Heating - Shadowing> | Ratio |
| (54509) YORP | 64.4 | 8.16 | 32.8 | 10 | 70 | 20 | 56.1 | 55.5 | 57.8 | 53.9 | 0.489 | -0.451 | -0.921 |
| (66391) 1999 KW4a | 487 | 103 | 171 | 0 | 80 | 20 | 55.5 | 43.7 | 48.0, 62.2 | 41.1, 70.6 | 0.015 | -0.020 | -1.379 |
| (66391) 1999 KW4b | 6.72 | 3.79 | 3.31 | 20 | 40 | 40 | 62.1 | 62.8 | 63.1 | 62.8 | 0.007 | -0.003 | -0.488 |
| (136617) 1994 CC | 367 | 5.36 | 11.3 | 0 | 80 | 20 | 56.0 | 62.3 | 61.1 | 62.3 | 0.002 | -0.002 | -0.944 |
| (276049) 2002 CE26 | 57.0 | 1.53 | 6.93 | 0 | 100 | 0 | 45.6 | 46.2 | 46.6 | 45.9 | 0.004 | -0.005 | -1.186 |
| 1998 KY26 | 36.0 | 4.53 | 10.4 | 0 | 90 | 10 | 58.8 | 58.1 | 60.1 | 58.7 | 0.065 | -0.078 | -1.203 |
| 2008 EV5 | 63.6 | 4.22 | 3.62 | 0 | 50 | 50 | 58.2 | 58.5 | 58.5 | 58.9 | 0.014 | -0.008 | -0.605 |
| Gaussian-Spheres** | 40.5 +80.2 / -28.7 | 4.85 +14.8 / -3.22 | 6.91 +20.0 / -3.91 | 0 +10 / -0 | 70 +30 / -40 | 30 +20 / -30 | 56.4 +3.3 / -5.5 | 56.2 +3.4 / -5.2 | 55.3 +4.2 / -8.9 | 56.2 +3.1 / -5.9 | -0.027 +0.157 / -0.120 | 0.027 +0.115 / -0.141 | -0.820 +0.406 / -0.593 |
| **Average** | 90.7 | 14.2 | 23.3 | 5 | 66 | 29 | 56.4 | 55.5 | 52.9 | 54.7 | N/A | N/A | -0.782 |
| **Best Match Frequency** | 6 | 90 | 28 | 5 | 94 | 25 | 13 | 93 | 15 | 3*** | N/A | N/A | N/A |

**Median and 1-$\sigma$ spread. ***Number of test asteroids in which neither of the 'convex', 'pseudo-convex', and 'shadowing' predictions match the 'self-heating' critical angle predictions.

**Figure Captions**

Figure 1: Schematic of the Yarkovsky and YORP effects on the orbit and spin properties of a small asteroid (copied from Rozitis & Green 2012).

Figure 2: Schematic of global self-heating occurring inside a large concavity of an asteroid.

Figure 3: Schematic of the ATPM where the terms $F_{\text{SUN}}$, $F_{\text{SCAT}}$, $F_{\text{RAD}}$, $k_c(dT/dx)$, and $\varepsilon\sigma T^4$ are the direct sunlight, multiple-scattered sunlight, re-absorbed thermal radiation, conducted heat, and thermal radiation lost to space, respectively (copied from Rozitis & Green 2011).

Figure 4: Asteroid (6489) Golevka radar-derived concave shape model (Hudson et al. 2000) and total view factor distribution. The green line corresponds to Golevka's north pole. The view on the left is a side view whilst that on the right is a view of Golevka's south pole.

Figure 5: Physical and shape properties for the test asteroids used. (a) Convex to concave volume ratio as a function of rotation period for the real asteroid shapes split into different rotation period groups. (b) Mean total view factor as a function of convex to concave volume ratio for both the real asteroid shapes and the Gaussian-spheres. The line is the best linear fit to the trend indicated, and the equation and $R^2$ value of the fit is given next to it. (c) Shape model frequency distributions as functions of convex to concave volume ratio for both the real asteroid shapes and the Gaussian-spheres. (d) Shape model frequency distributions as functions of YORP-coefficient for both the real asteroid shapes and the Gaussian-spheres. The YORP-coefficient values used here are calculated assuming 0° obliquity and no global self-heating effects.

Figure 6: Mean percentage difference of the 'convex', 'pseudo-convex', and 'self-heating' (legend) orbital drift predictions relative to the 'shadowing' predictions as a function of obliquity (x-axis) for: (a) the diurnal Yarkovsky effect, and (b) the seasonal Yarkovsky effect. The error-bars on the 'self-heating' predictions represent the 1-sigma ranges of difference variations caused by the range of test asteroids used

Figure 7: YORP-rotational-acceleration acting on all real asteroid shapes used as a function of obliquity (x-axis) and model variant (legend).

Figure 8: YORP-function difference for the 'convex' (left panels), 'pseudo-convex' (middle panels), and 'shadowing' (right panels) model variants as functions of convex to concave volume ratio (top panels) and 0° obliquity 'self-heating' YORP-coefficient (bottom panels). All test asteroid shapes used are plotted (legend of top left panel), and the median values of the different model variants are also indicated by the horizontal lines (legend of bottom left panel).

Figure 9: Comparison of the 'convex' (left panel), 'pseudo-convex' (middle panel), and 'shadowing' (right panel) critical angles with the 'self-heating' critical angles (x-axis). The diagonal lines represent the trend if the two compared values were exactly equal to one another. All test asteroid shapes used are plotted (legend of middle panel).

Figure 10: YORP-function vertical offset ratios, <self-heating-shadowing>/<shadowing-pseudo-convex>, as functions of volume ratio (left panel) and YORP-coefficient (right panel). All test asteroid shapes used are plotted (legend of left panel), and the median value is also indicated by the horizontal line.

Figure 11: YORP-rotational-acceleration acting on four extreme case asteroids as a function of obliquity and model variant. The red and blue lines represent model predictions with and without global self-heating effects included respectively. The solid, dashed-dotted, and dotted lines represent model predictions with smooth, 50% rough, and 100% rough surfaces respectively. The error-bars represent the 1-sigma uncertainty on the 50% rough surface predictions when the degree of surface roughness is allowed to vary in a patchy way across the surface.

Figure A1: The diurnal and seasonal Yarkovsky-orbital-drifts acting on asteroid (1620) Geographos as a function of obliquity. The seasonal effect has been multiplied by a factor of -1 so that it can be plotted on the same axes as the diurnal effect. This is a corrected version of figure 7c given in Rozitis & Green (2012).

**Figures**

Figure 1:

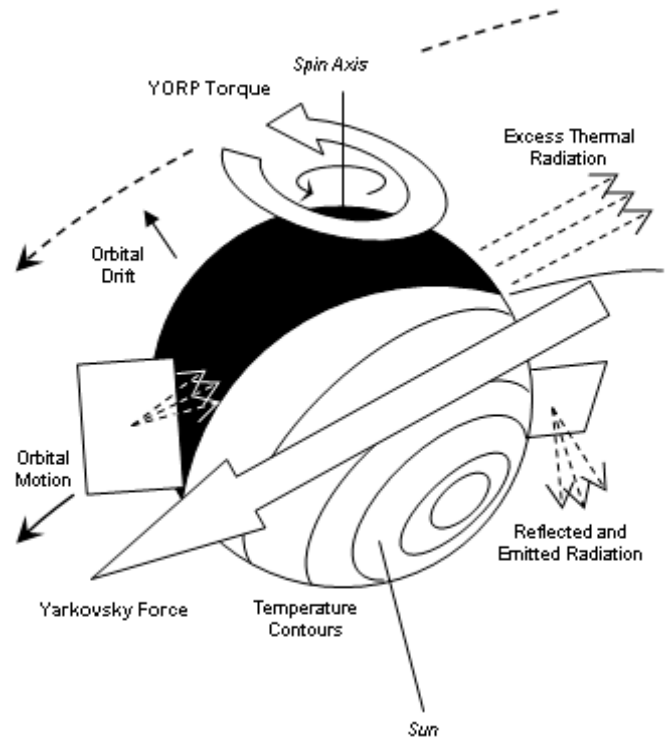

Figure 2:

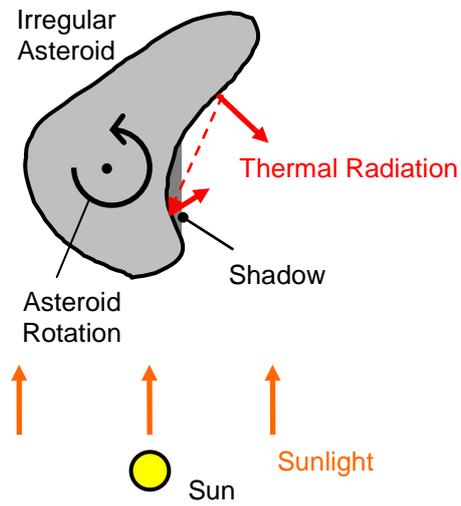

Figure 3:

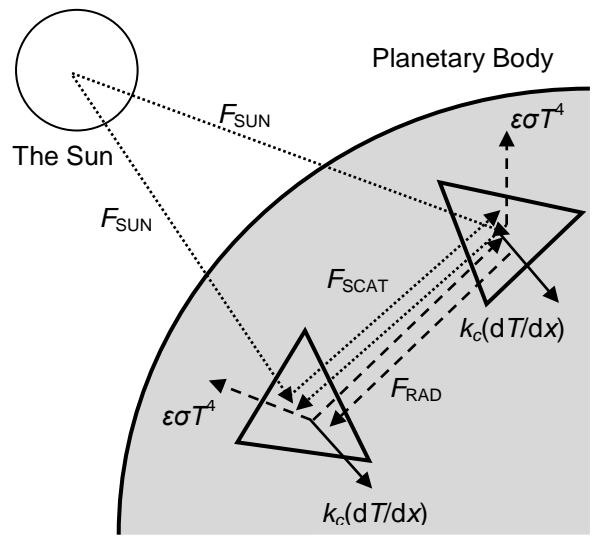

Figure 4:

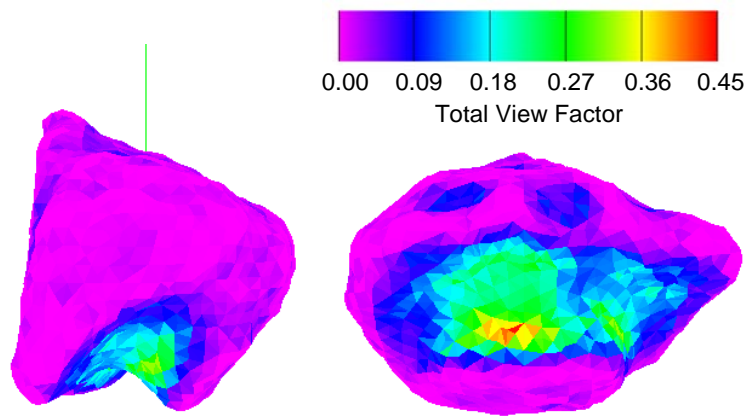

Figure 5:

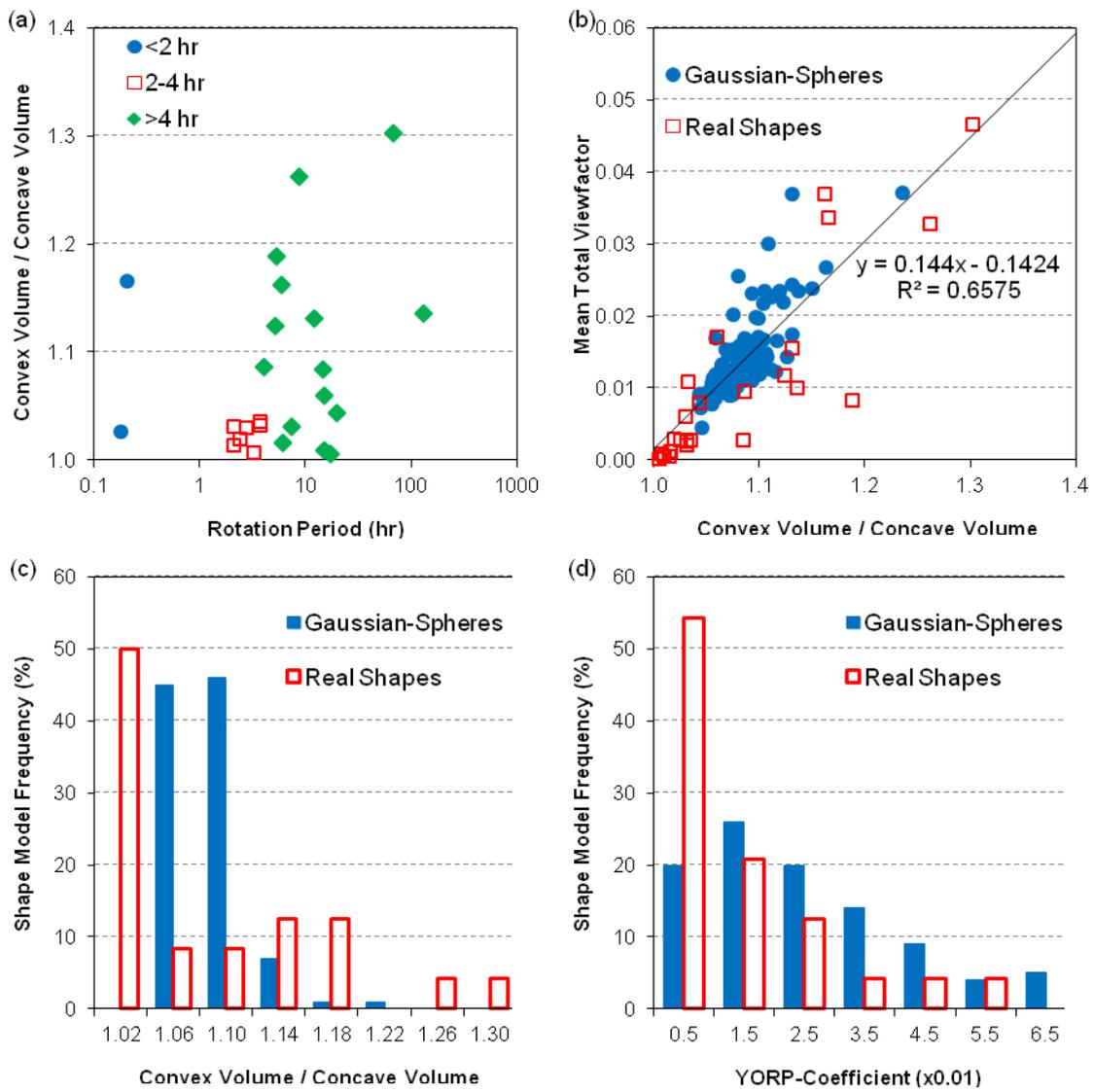

Figure 6:

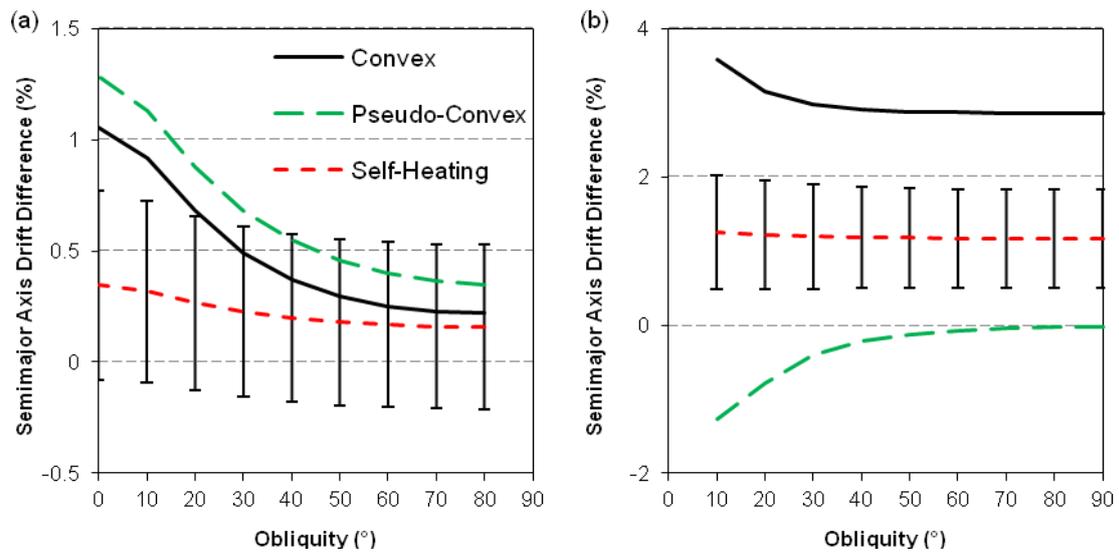

Figure 7:

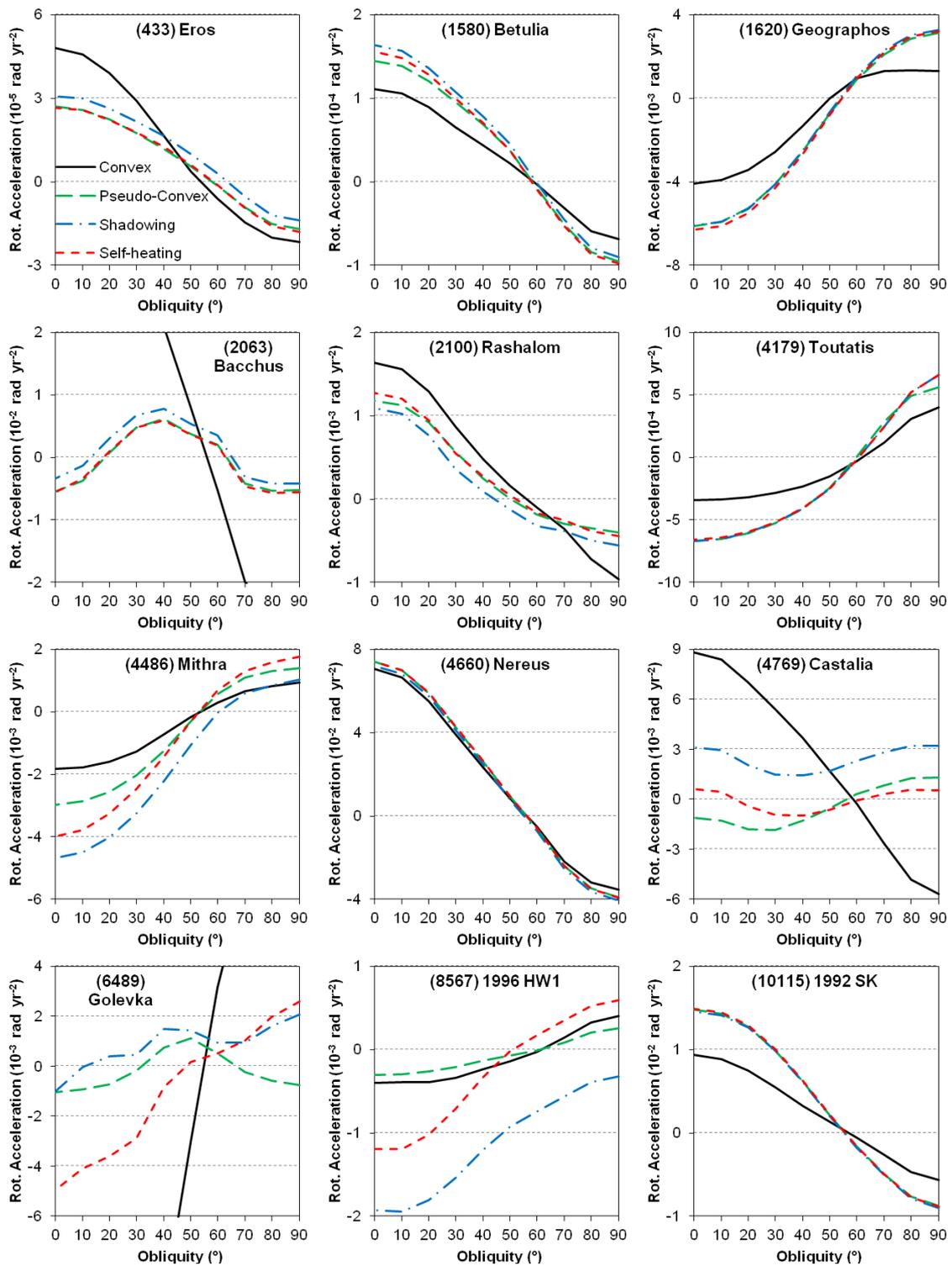

Figure 7 (continued):

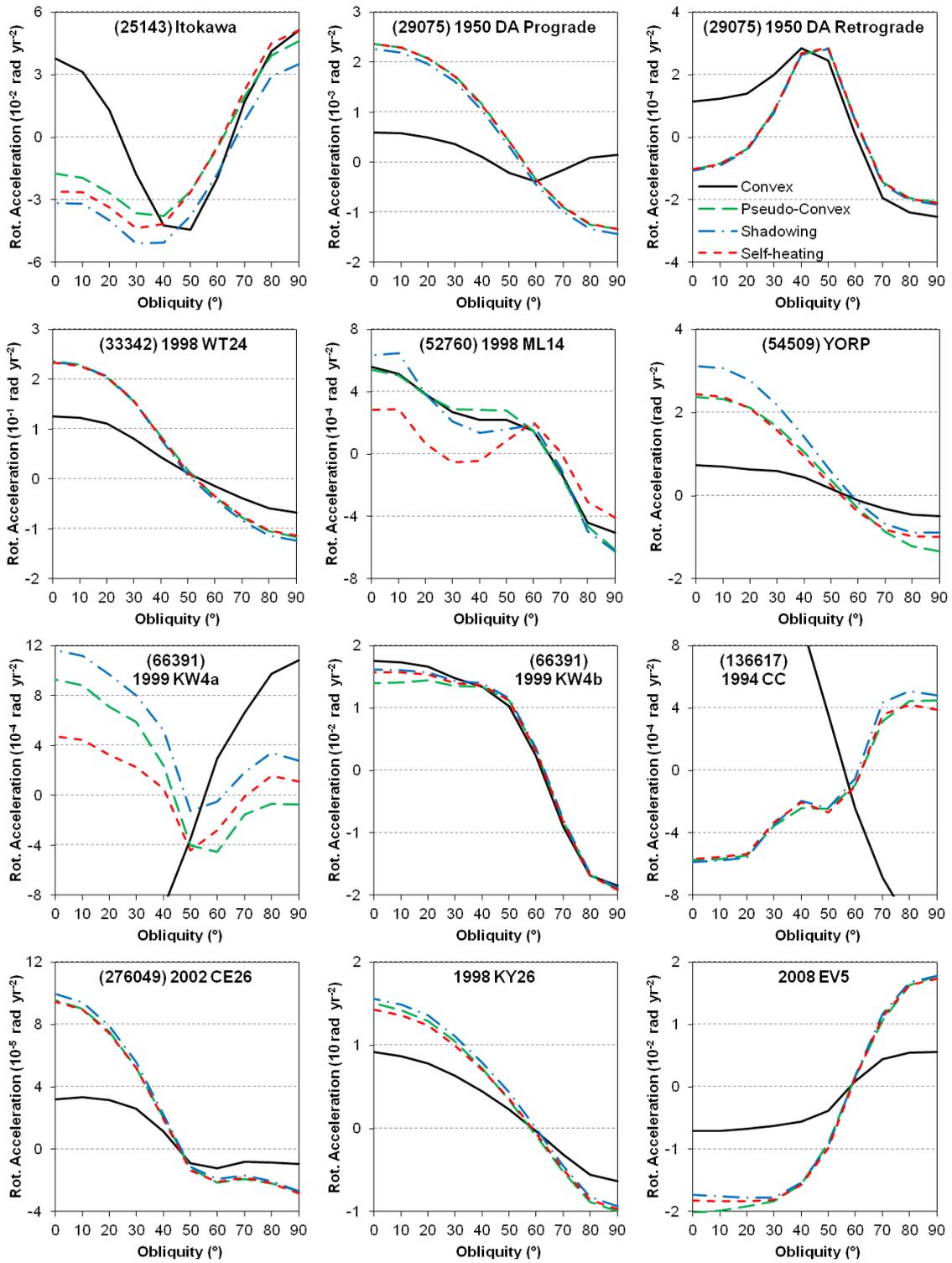

Figure 8:

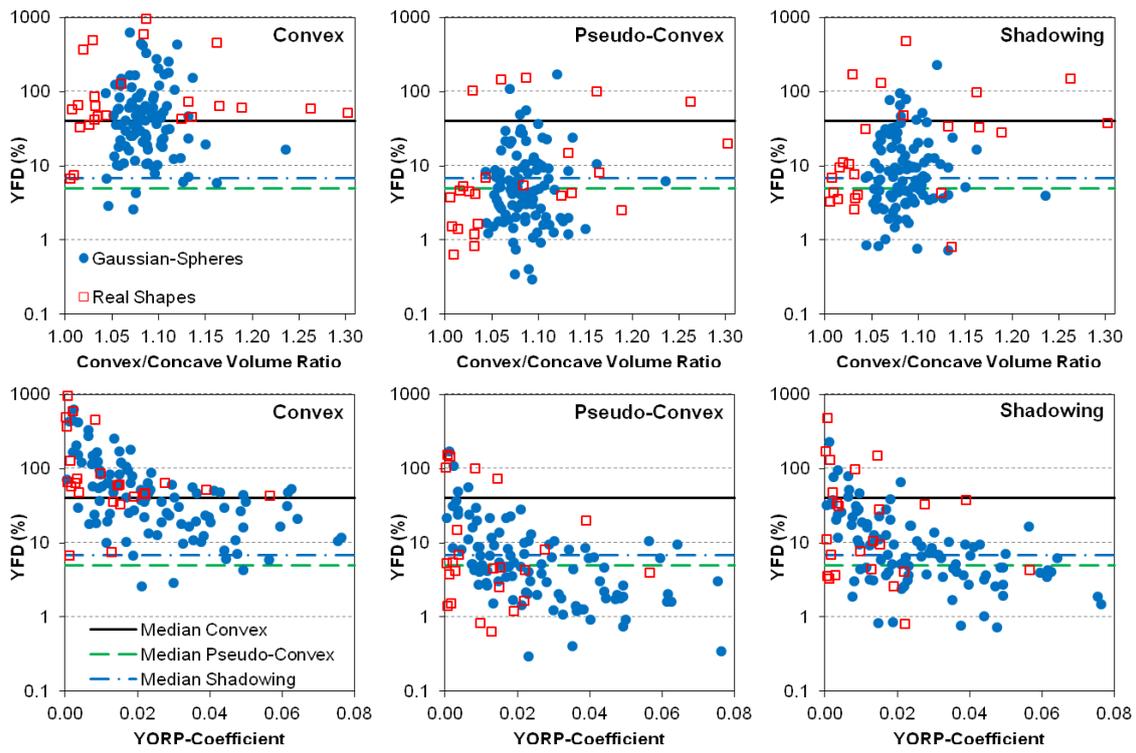

Figure 9:

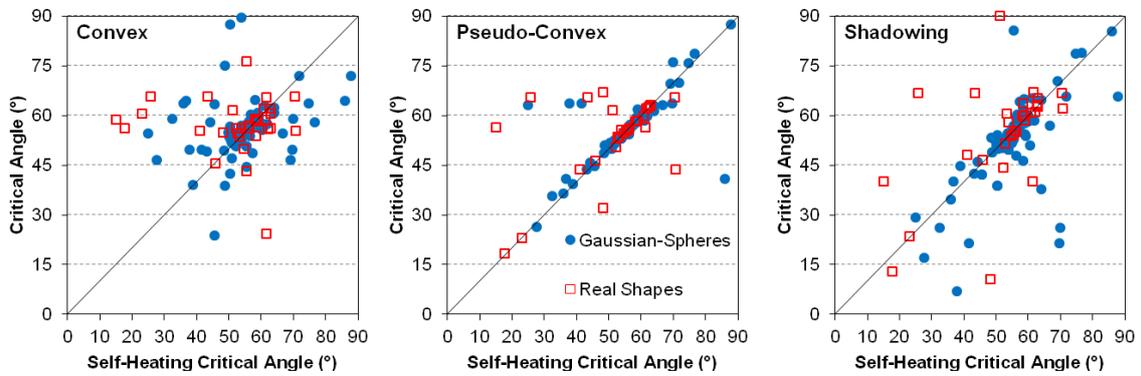

Figure 10:

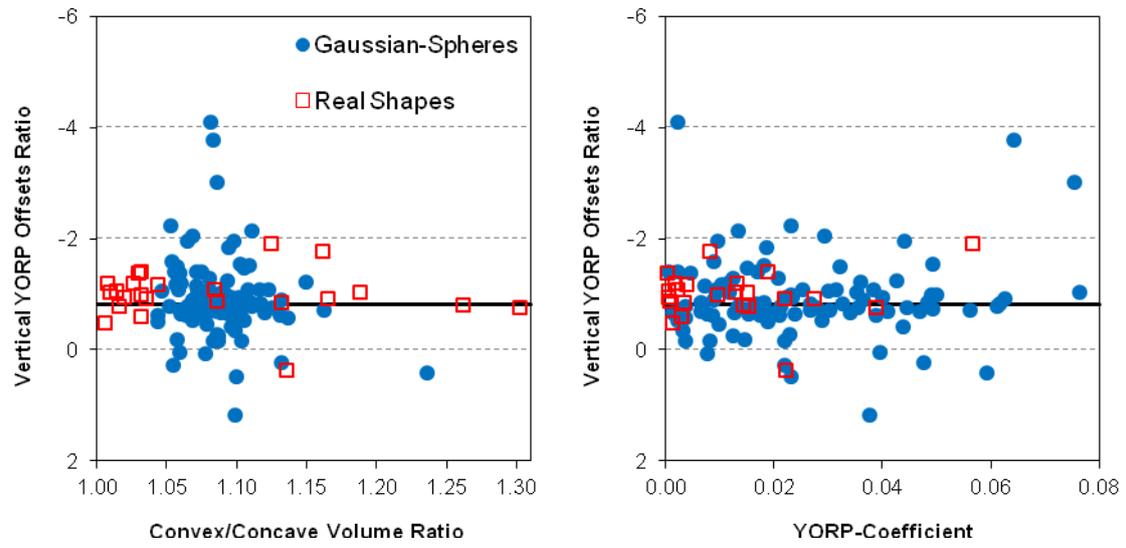

Figure 11:

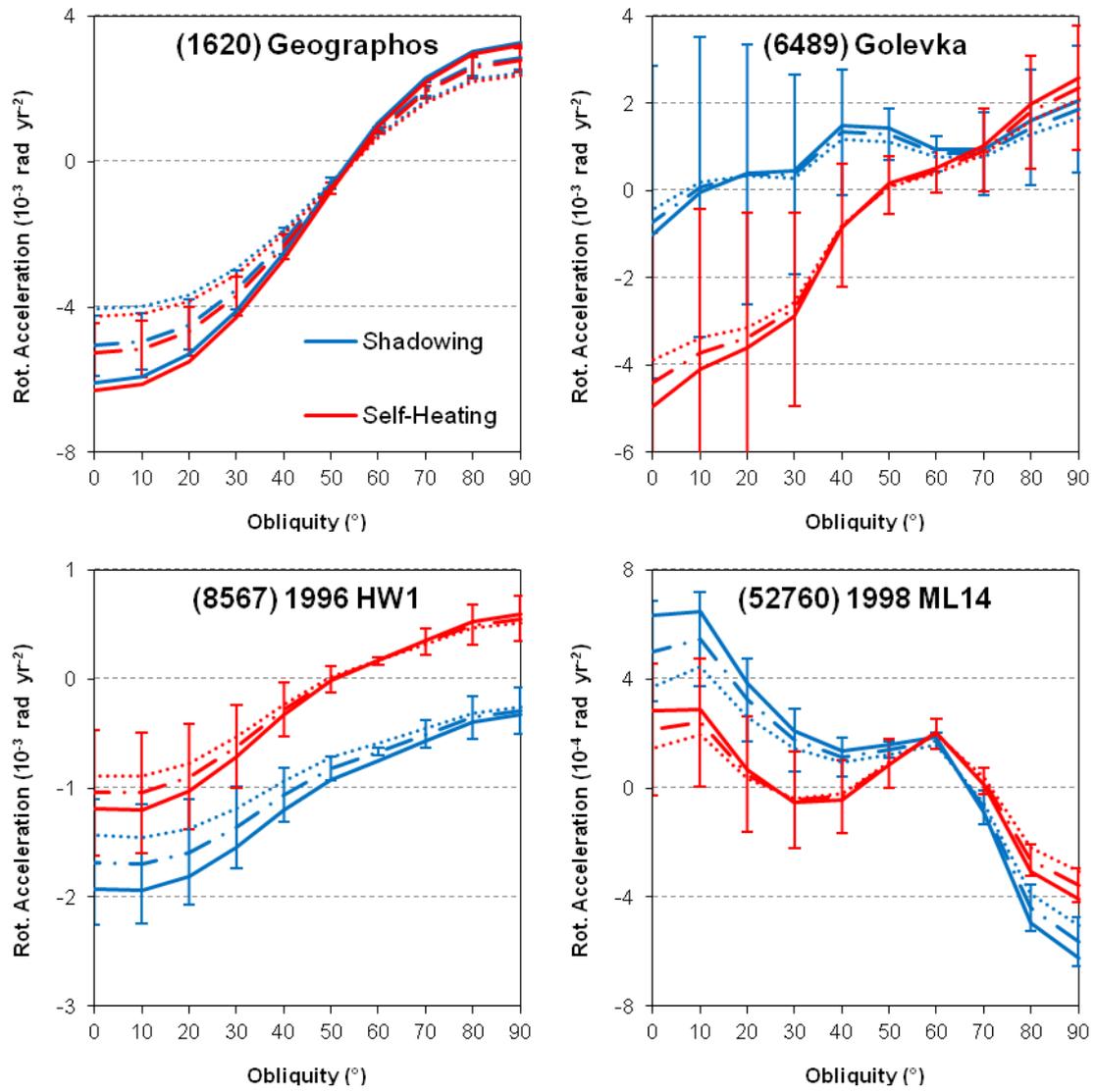

Figure A1:

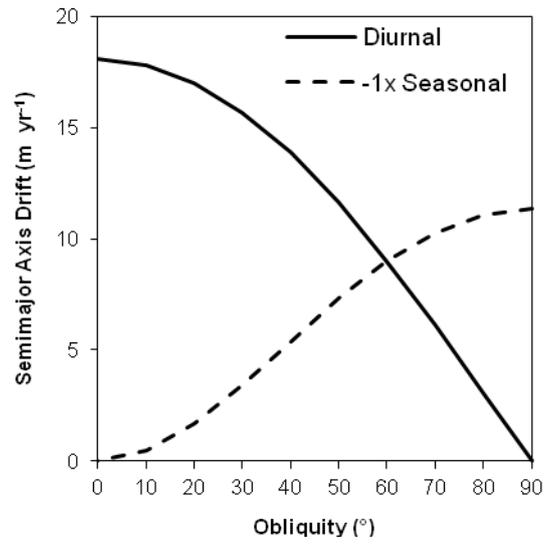